\documentclass[aps,prx,showpacs,superscriptaddress,groupedaddress]{revtex4-2}
\usepackage{xr}
\usepackage[utf8]{inputenc}
\usepackage[T1]{fontenc}
\usepackage{amsmath}
\usepackage{amsfonts}
\usepackage{amssymb}
\usepackage{graphicx}
\usepackage{lineno}
\usepackage{xcolor}
\usepackage{hyperref}

\usepackage{graphicx} 
\usepackage{color}
\usepackage{amsmath}

\begin{document}
\title{Dominance to egalitarian transition in diverse communities}
\author{David A. Kessler and Nadav M. Shnerb}
\date{June 2024}
\begin{abstract}
\noindent Diverse communities of competing species are generally characterized by substantial niche overlap and strongly stochastic dynamics. Abundance fluctuations are proportional to population size, so the dynamics of rare populations is slower. Hence, once a population becomes rare, its abundance gets stuck at low values.   Here, we analyze the effect of this phenomenon on community structure. We identify two different phases: a dominance phase, in which a tiny number of species constitute most of the community, and an egalitarian phase,  where it takes a finite fraction of all species to constitute most of the community.  We demonstrate the validity of the theory using empirical findings for a variety of hyperdiverse communities, and clarify the role of demographic stochasticity in shaping patterns of commonness and rarity.
\end{abstract}
\maketitle

\maketitle

\section{Introduction}

Many ecological systems contain numerous competing species, strains or types~\cite{plankton,stomp2011large,ter2013hyperdominance,connolly2014commonness,fierer2007metagenomic}. In such systems, it can be assumed a priori that the structure of the community essentially reflects resource competition among species~\cite{gause2003struggle,tilman1982resource}, with factors such as niche overlap and fitness differences playing a central role in the community assembly~\cite{chesson2000mechanisms,chesson2003quantifying}. Analyzing these factors and their impact is crucial for understanding the dynamics of these systems, and consequently, for our ability to intervene in these dynamics to achieve desired outcomes—from maintaining biodiversity in a changing world to successfully altering the state of the gut microbiome~\cite{david2014host,grilli2020macroecological, eguiluz2019scaling, cooper2024consistent, callaghan2021global}. 

Unfortunately, progress in this area  has been quite difficult. The coexistence of many species remains largely puzzling, especially given the competitive exclusion principle~\cite{tilman1982resource} and the strict constraints on the stability and feasibility of such complex systems~\cite{may1972will}. Moreover, quantifying the relevant parameters in diverse communities is extremely challenging, particularly considering the high level of stochasticity typically present in ecological dynamics. As a result, approaches inspired by statistical physics, which examine generic models through a few summary statistics, have gained significant popularity in recent years~\cite{fisher2014transition,kessler2015generalized,bunin2017ecological,barbier2018generic,grilli2020macroecological,van2024tiny}.

Broadly speaking, attempts to present a generic analysis of diverse communities can be divided into two main approaches. One approach assumes that interspecific interactions are much weaker than intraspecific interactions, so the niche overlap between any pair of species is relatively small~\cite{fisher2014transition,bunin2017ecological,barbier2018generic,azaele2023large,marcus2022local}. The alternative approach assumes high niche overlap but small fitness differences; the simplest examples of this approach are what are known as neutral models.

In the original neutral models proposed by Kimura \cite{crow1970introduction} and Hubbell \cite{hubbell_book}, demographic stochasticity -- namely, the inherent randomness in the birth-death process at the individual level -- is the sole driver of abundance variations. Analytical solutions to these models are relatively easy to obtain \cite{maritan1, azaele2015towards} and have been quite successful in explaining the observed species abundance distributions (SAD) in both regional and local communities \cite{TREE2011}, as well as other "static" patterns. However, pure demographic stochasticity cannot account for dynamic patterns, whether evolutionary (such as the time to the most recent common ancestor) \cite{nee2005neutral, ricklefs2006unified} or ecological (such as the dynamics of abundance variation and similarity indices) \cite{leigh2007neutral, chisholm2014species, kalyuzhny2014niche, kalyuzhny2014temporal}. Demographic stochasticity results in relatively slow and weak abundance variations, whereas observed variability is much stronger and occurs more rapidly \cite{kalyuzhny2014niche, chisholm2014temporal}.

The time-averaged neutral model~\cite{kalyuzhny2015neutral} addresses these limitations by relaxing the assumption of time-independent fitness. In this model, the relative fitness of each species fluctuates over time, but all species share the same mean fitness. The analysis of the time-averaged neutral model is more complex, as environmental stochasticity can stabilize coexistence through the storage effect~\cite{chesson1981environmental}. However, the significance of this effect diminishes in highly diverse systems~\cite{dean2017fluctuating,danino2018theory,pande2022temporal,meyer2022storage}.

Recent studies have highlighted another notable effect of environmental stochasticity, known as "stickiness"~\cite{van2024tiny} or "diffusive trapping"~\cite{dean2020stochasticity}. Under environmental stochasticity, abundance variations are proportional to population size, meaning that the dynamics of rare species is slow, causing them to linger near the extinction point for extended periods. \citet{van2024tiny} suggested that this stickiness enables time-averaged neutral models to produce patterns -- such as changes in abundance over time, dominant species turnover, and community evenness -- that closely resemble those observed in real ecological communities. Similar results were found by \citet{mallmin2024chaotic} and his colleagues in a system of competing species for which the overall dynamics are chaotic. This makes sense, because the relative fitness of a specific species depends on the abundances of its competitors, and if these abundances fluctuate strongly over time, it is likely that the community will reach a state of time-averaged neutrality.

Our main goal in this paper is to point out the existence of a previously unknown phase transition between dominance communities and egalitarian communities. To describe this phase transition, we will adhere to the commonly accepted definition in the literature, according to which the number of dominant species, $S_{1/2}$, is the minimum number of species that must be grouped together at a given moment to account for more than half of the total population of the community. The level of equality in the community is measured by the evenness ratio $S_{1/2}/S$, where $S$ is the total species richness. We will show that in the dominance phase, the growth of $S_{1/2}$ with $S$ is sub-linear, so the evenness parameter approaches zero as $S$ tends to infinity, whereas in the egalitarian phase, $S_{1/2}$ is linear in $S$, and therefore the evenness parameter reaches a finite value. We then demonstrate these two phases through the analysis of empirical data on commonness and rarity patterns across a wide range of ecological systems, from trees in tropical forests to the human microbiome.

During the analysis and comparisons with the empirical data, we also arrived at two very important insights.

\begin{itemize} 

\item The basic model, as presented by \citet{van2024tiny}, includes dynamics of growth, competition, and environmental stochasticity only. Immigration was presented, in one of the supplementary discussions, only as an additional option. We will show that without a mechanism like immigration, which imposes a lower bound on the abundance of each species, the community reaches a state of monodominance, where all populations except one decline below any finite abundance level. Therefore, the process of immigration (that was considered in one of the supplementary of ~\cite{van2024tiny}) is a crucial factor that must be included in the analysis, and small changes in the immigration rate can change the phase in which the community resides.

\item Moreover, the basic model of \cite{van2024tiny} includes only environmental stochasticity and does not take demographic stochasticity into account. Such an approximation -- neglecting demographic stochasticity -- can only be made for communities with enormous numbers of individuals. We will show that for populations of microorganisms, where the number of individuals is indeed vast, the system can be described by environmental stochasticity alone. However, for macroorganisms, the abundance distribution of relatively rare species is significantly affected by demographic stochasticity, and to properly describe it, this mechanism must also be taken into account.
\end{itemize}

This phase transition has practical implications. A small decrease in the immigration rate or increase in the level of stochasticity can lead to a sharp decline of the diversity within a given community. Moreover, it is likely that this transition is not a specific feature of the time-averaged neutral model but will appear generically in systems with strong interactions (high niche overlap) that support chaotic dynamics. We will discuss these topics further in the last section below.

\section{Materials and methods}

\noindent {\bf Data Compilation and Analysis}: As explained below, the phase transition we analyze in this paper manifests itself in two characteristics: the species abundance distribution (SAD) and the evenness parameter $S_{1/2}/S$. Like any phase transition in complex systems, the distinction between the two phases becomes sharper as the system grows larger, so that $S$ increases. To test our predictions, we used several recent large databases on communities with thousands of species: human microbiome~\cite{david2014host}, marine prokaryotes~\cite{eguiluz2019scaling}, tropical forest trees~\cite{cooper2024consistent}, and birds (worldwide)~\cite{callaghan2021global}. \\

\noindent  {\bf The neutral model}: We consider a community of $S$ populations, with differential responses to environmental variations. $N_i$ is the abundance of the $i$-th species, and the  carrying capacity parameter is denoted by $K$. With no stochasticity and no immigration, $N_i$ satisfies the equation,
\begin{equation}  \label{eq1}
   \frac{ dN_i}{dt} =  N_i \left( 1-\frac{\sum_{j=1}^S N_j}{K} \right). 
\end{equation}
As all species admit the same growth rate and interact in a symmetric manner, the model is neutral. In particular, any solution that satisfies $\sum_{j=1}^S N_j=K$ is a steady state of the system.\\

\noindent  {\bf The time-averaged neutral model}: To allow for nontrivial dynamics, one would like to add environmental stochasticity and immigration to the process described in~\ref{eq1}. The rate of immigration is denoted by $\mu$, and the growth rate of each species undergoes an Ornstein–Uhlenbeck process with a correlation time $\tau$. The corresponding set of stochastic differential equations is, 
\begin{align} \label{eq2}
   \frac{ dN_i}{dt}  = \mu + N_i &\left( 1-\frac{\sum_{j=1}^S N_j}{K} \right) dt + r_i(t) N_i \nonumber \\
   \frac{ dr_i}{dt}  &= -\frac{r_i}{\tau} + \theta \eta_i(t), 
\end{align}
where $\eta_i(t)$ is a white-noise process. The index $i$ of $\eta_i$ indicates that each of the $S$ species responds to the environmental variations in an independent manner. The strength of environmental stochasticity is given by ${\tilde \sigma}_e^2 = \frac{\theta^2 \tau}{2}$. \\

\noindent  {\bf Fokker-Planck (FP) equations}:The main analytical results presented in this paper are based on solving Fokker-Planck equations. To derive these equations, we assumed a relatively short correlation time $\tau$ for the environmental stochasticity, which allows the use of a single effective equation for the process described in Equation \ref{eq1}. The relevant considerations are detailed in SI \ref{suppD}.\\

\noindent  {\bf Demographic stochasticity}: refers to the intrinsic randomness of the birth-death process at the individual level. This stochasticity also manifests in erratic abundance variations, but the intensity of these fluctuations is weaker (compared to environmental stochasticity) when the population is large, and therefore it was not included in Eq. \ref{eq1} above  or in the analyses of \cite{van2024tiny} and \cite{mallmin2024chaotic}. However, demographic stochasticity is crucially important in small populations and in extinction processes; as we shall see, it must be considered outside the microorganism realm.  To include demographic stochasticity in the treatment, one must add a term to the relevant equation that expresses noise whose amplitude scales with the square root of the population size, as opposed to environmental stochasticity, whose amplitude scales linearly with population size. The technical details can be found in SI \ref{SuppE}.\\

\section{Results}

As we have already mentioned, without external immigration, the stickiness will eventually cause the abundance of all species, except one, to drop below any finite threshold, leading the community to a state of monodominance. To avoid disrupting the continuity of the main discussion, we detail and demonstrate this claim in SI \ref{suppC}. Below, we assume $\mu>0$ and examine the behavior of the system given this assumption.

\subsection{Stickiness and Species Abundance Distribution}

Without immigration, the carrying capacity parameter \(K\) plays no role in the dynamics. Actually, by defining \(N_i = K n_i\) one may rescale Eq. (\ref{eq1}) to  \(K=1\). When immigration is introduced, the parameter $K/\mu$ expresses the ratio between the carrying capacity and the minimum level of population size. The larger this ratio, the stronger the stickiness effect.

On top of that, since stickiness arises from environmental stochasticity, it becomes stronger as \({\tilde \sigma}_e^2\) increases. Dimensional analysis reveals that the dimensionless parameter that governs stickiness is $$\gamma \equiv \frac{K {\tilde \sigma}_e^2}{\mu}.$$ The larger $\gamma$ is, the stronger is the stickiness, and the community approaches monodominance  when $\gamma \to \infty$, e.g., when $\mu \to 0$. 

In order to advance in the analysis, we are interested in replacing Equation (\ref{eq2}), which provides us with a description of an $S$-dimensional system where each species can affect every other species, with an effective, one-dimensional equation for a focal species, considering all others as a single rival species.  In the classical neutral theory, with pure demographic stochasticity, this can be done trivially, as the species identity of a particular individual plays no role in the dynamics. In the time-averaged neutral theory, however, the situation is much more subtle.

The distinction between neutral and time-averaged neutral models was clarified in \cite{danino2018theory} and \cite{steinmetz2020intraspecific}, and has to do with the distinction between the mean (over time) of the growth rate of a species and the the mean (over individuals) growth rate of the community. In Eq. (2)  the linear growth rate of each species is $1+r(t)$. Since the $r$ process is symmetric around zero,  the time-average growth rate of each species is unity. Nevertheless, the instantaneous growth rate of the community is, on average, greater than one.  At any given moment, the fitter species are growing faster, so on average more than $50 \%$ of the individuals belong to instantaneously  beneficial species. As a result, the typical value of $\sum_j N_j >K$, and therefore the dynamics of a single species satisfies the effective one-dimensional stochastic differential equation, 
\begin{equation} \label{eq4}
    \frac{dN}{dt} = \mu +\left(\frac{ \sigma_e^2}{2}-d\right) N +  \sigma_e \eta(t) N, 
\end{equation}
where $d = \mathrm{E} [(\sum_j N_j -K)/K]$, $\sigma_e$ reflects the combined effect of the environmental variations that act directly on that given species (denoted ${\tilde \sigma}_e$)  and the fluctuations in $\sum_j N_j$. The Stratonovich term $ \sigma_e^2/2$,  expresses the fact that for a population that sometimes grows exponentially and sometimes declines, even if its average growth rate is zero, the arithmetic mean still increases over time. Once this term is introduced, the  standard Ito calculus may be applied to Eq. (\ref{eq4}). The two parameters $d$ and $ \sigma_e$  may be measured, for any given values of $K$, $S$ and $\sigma_d$, from long simulations of Eq. (\ref{eq2}).

\begin{figure}[hbt!]
	\centering{
		\includegraphics[width=8cm]{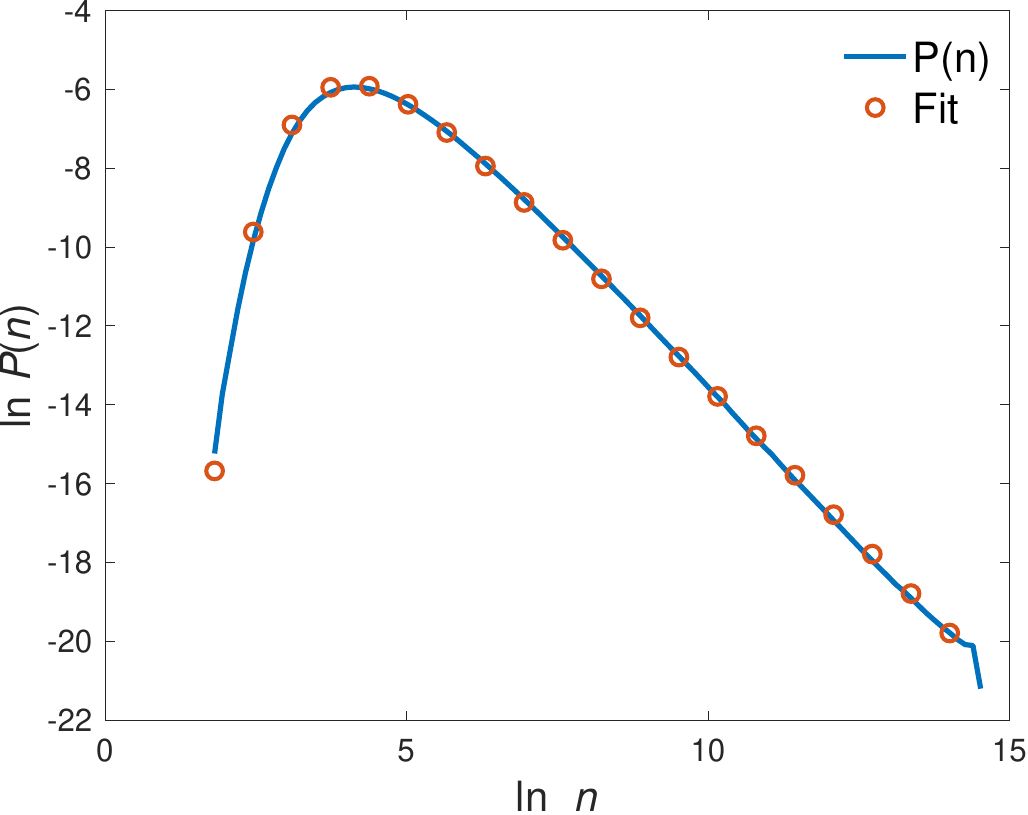}}
	\caption{Red circles: results of a simulation of the process described in Eq. (\ref{eq2}) with parameters $S= 200$, $\mu = 0.6$, $K=2 \cdot 10^6$, $\sigma_e = 0.025$ and $\tau =8$. The values of the extracted parameters for $\ref{eq5}$ are $d=0.0034$ and $ \sigma_e = 0.0134$.  The  full line in blue is the expression of  Eq. (\ref{eq5}) with parameters $\alpha = 91.2 $ and $\beta = 1.56$.    \label{fig1}}
\end{figure}

Once the relevant parameters are calibrated, the distribution for $P(N)$ may be extracted from Eq. (\ref{eq4}), see SI \ref{suppD}. The resulting distribution is, 
\begin{equation} \label{eq5}
P(n) = A e^{-\alpha/n} n^{-\beta},
\end{equation}
where $\alpha = 2\mu/\sigma_e^2$, $\beta = 1+2d/\sigma_e^2$ and $A$ is a normalization factor. Figure \ref{fig1} illustrates the success of the approximation and the validity of Eq. (\ref{eq5}) in a simulation of a community of $S=200$ interacting species governed by Eq. \eqref{eq2}.

A similar result was presented a few months ago by \citet{mallmin2024chaotic},  who dealt with a system of competing species  where the community dynamics is chaotic (but without external stochasticity). In such a case, one can consider, for each focal species, all other species as an effective external environment whose fluctuations generate stochasticity in the instantaneous growth rate. A discussion of the similarities and differences between our case and the chaotic model will be presented below.

\subsection{The egalitarian transition}

The most striking implication of Equation (\ref{eq5}) is the sharp shift in the compositional properties of the community at $\beta = 2$. When $\beta > 2$, abundant species are relatively rare, resulting in an ``egalitarian'' community. Conversely, when $\beta < 2$, the community composition is dominated by a few exceptionally abundant species.

To quantify the evenness of the community, one may use the criterion suggested by \citet{van2024tiny}, which involves comparing the total number of species, $S$, with the minimal number of species required to make up half of the community, $S_{1/2}$. In a more egalitarian community, the fraction $S_{1/2}/S$ is finite, indicating that the number of dominant species is proportional to the total number of species. In contrast, in a community dominated by only a few species, the ratio $S_{1/2}/S$ approaches zero as $S$ increases, meaning the number of dominant species is sublinear in $S$.

\begin{figure}[hbt!]
	\centering{
		\includegraphics[height=6cm]{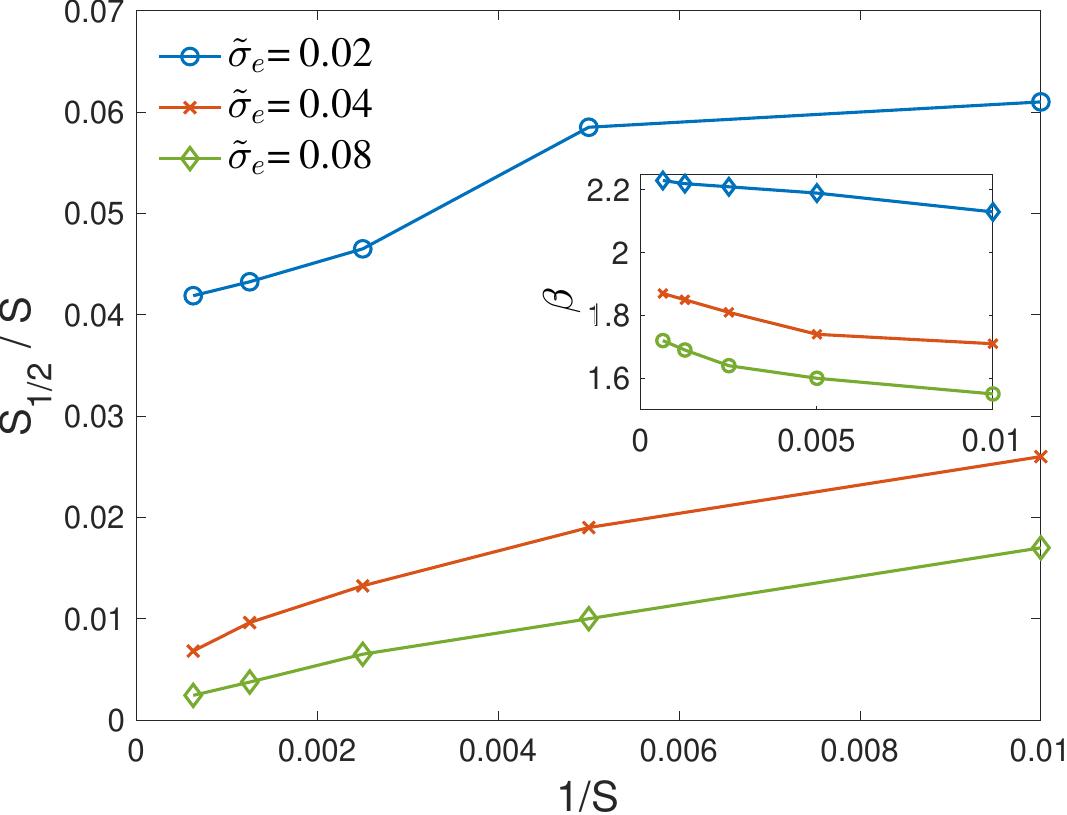}\includegraphics[height=6cm]{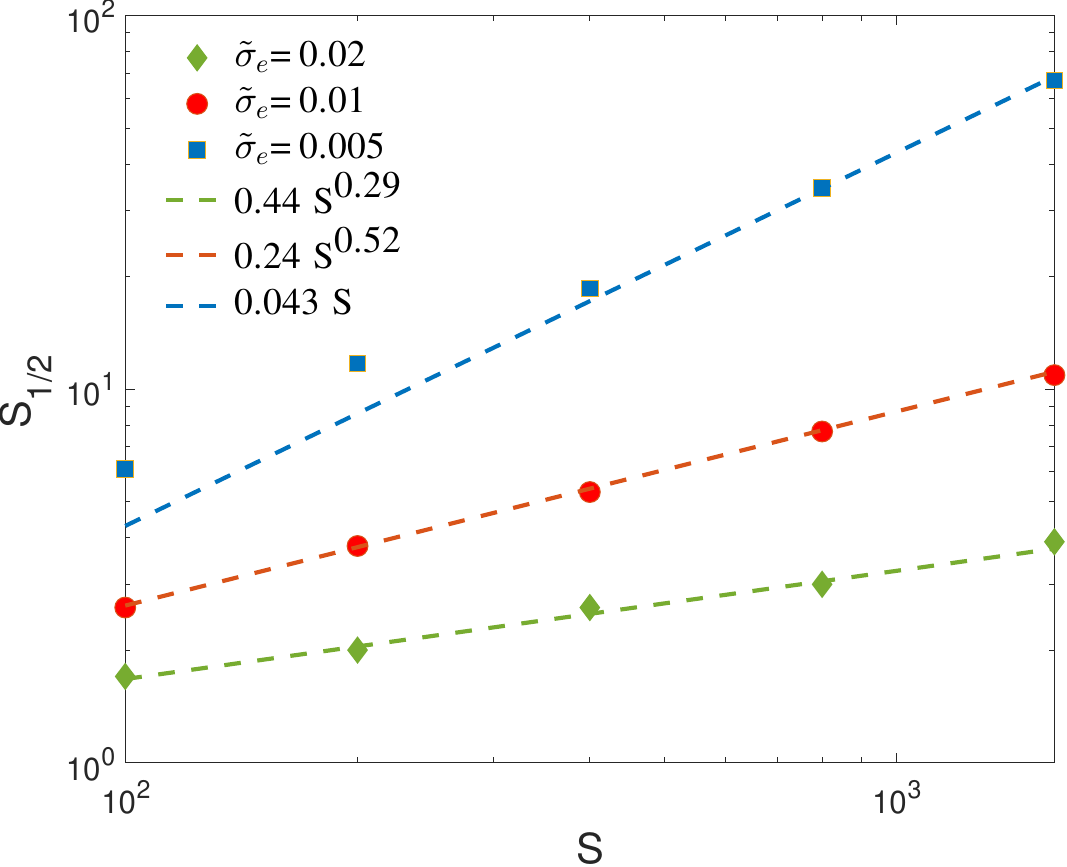}  }
	\caption{{\bf Panel A, left}: The evenness parameter $S_{1/2}/S$ is plotted against $1/S$, the total number of species, for ${ \tilde \sigma_e} = 0.005$  (blue) $0.01$ (red) and $0.02$ (green). For all these simulations $\mu = 0.6$, $\tau =8$ and $K = 2 \cdot 10^6$. The values of the parameter $\beta$, as extracted from the simulation, are plotted (again, vs. $1/S$) in the inset. As expected, for $\beta >2$ (blue) the evenness parameter $S_{1/2}/S$  extrapolates, at $S \to \infty$, to a finite value, meaning that the fraction of high-abundance species is finite and the community is egalitarian. This feature is also reflected in {\bf right panel, (B)}, where $S_{1/2} \sim 0.04 S$ at large $S$.   On the other hand, when   ${ \tilde \sigma_e} = 0.02$ (green), the value of $\beta$ approaches $\sim 1.8$ and is definitely smaller than two. Concurrently, the evenness parameter extrapolates to zero at large $S$, meaning that the community is dominated by a small number of species. The results shown in panel (B) suggest $S_{1/2} \sim S^{0.3}$. The case  ${ \tilde \sigma_e} = 0.01$ (red) represents a near marginal case, for which the $\beta$ extrapolates to values slightly smaller than $2$, and it would appear that $S_{1/2}/S$ is tending toward zero. Indeed, panel (B) shows a power law with exponent close to $1/2$.      \label{fig2}}
\end{figure}

In Appendix \ref{suppE}, we provide the relevant mathematical analysis, demonstrating that, as $\beta$ decreases towards 2,  $S_{1/2}/S$ monotonically decreases, and in the limit  $\beta \to 2^+$,
\begin{equation}
\frac{S_{1/2}}{S} \sim e^{-(\ln 2)/(\beta-2)}.
\end{equation}
Thus, this evenness parameter vanishes in a singular manner in that limit, indicating the transition of out the egalitarian phase. 

The numerical results presented in Figure \ref{fig2}(A) illustrate this phenomenon and offer several additional interesting insights. Beyond the positive indication regarding the result in the limit where $S \to \infty$, Figure \ref{fig2}(A) shows that the value of 
$\beta$ depends only weakly on $S$ so it can be considered approximately constant. Moreover, as seen in Figure \ref{fig2}(B), for $\beta<2$ the dependence of $S_{1/2}$ on $S$ follows a power law, with an exponent approaching unity at the phase transition point $\beta=2$.

\subsection{Comparison with Empirical Results and the Impact of Demographic Stochasticity}

Let us now examine the species abundance distribution in several cases of diverse communities, assess their degree of alignment with the results presented above, and attempt to differentiate between dominance and egalitarian communities, linking the results to the fundamental characteristics of each system.

We analyze the community structure in four systems: the human gut microbiome~\cite{david2014host}, marine prokaryotes~\cite{eguiluz2019scaling}, tropical trees~\cite{cooper2024consistent}, and bird species~\cite{callaghan2021global}. All of these communities are hyperdiverse, with thousands of species, making them reflective of the limit where $1/S$ is very close to zero -- a limit where the distinction between egalitarian and dominance systems is clearly defined, with the relevant values of $S_{1/2}/S$? being those shown in Figure \ref{fig2} above.

\begin{figure}[hbt!]
	\centering{
		\includegraphics[width=8cm]{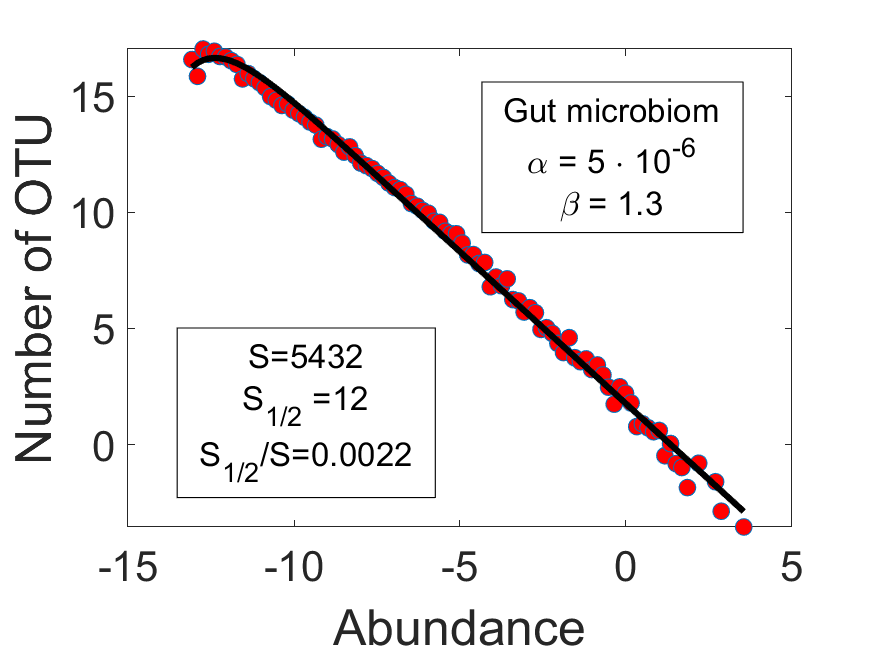} \includegraphics[width=8cm]{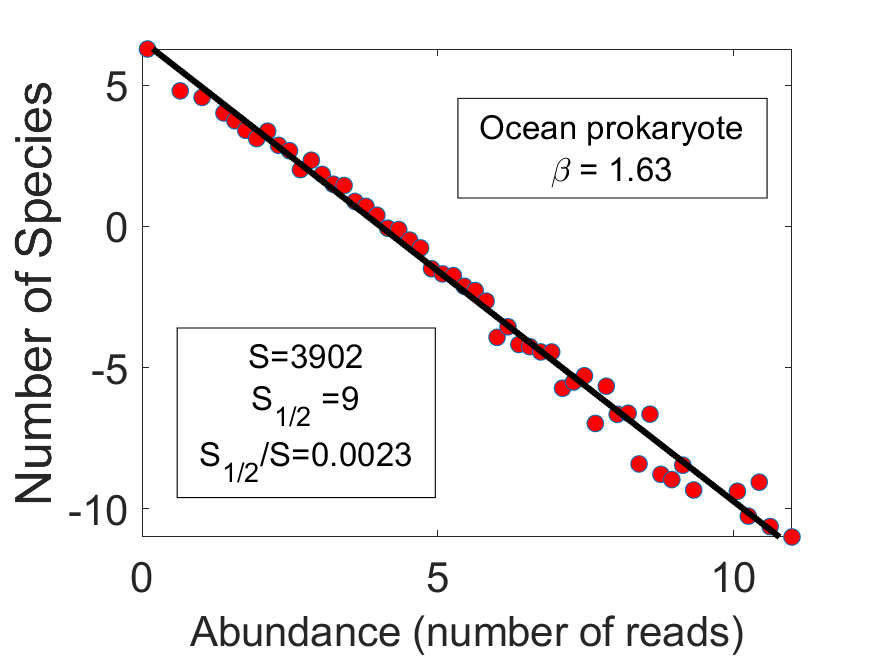} \\ \includegraphics[width=8cm]{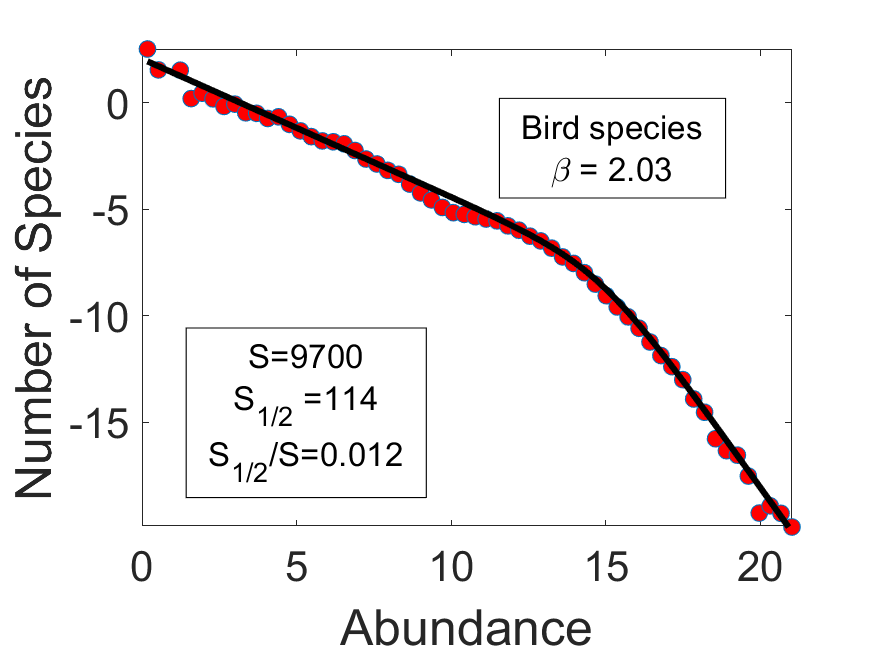} \includegraphics[width=8cm]{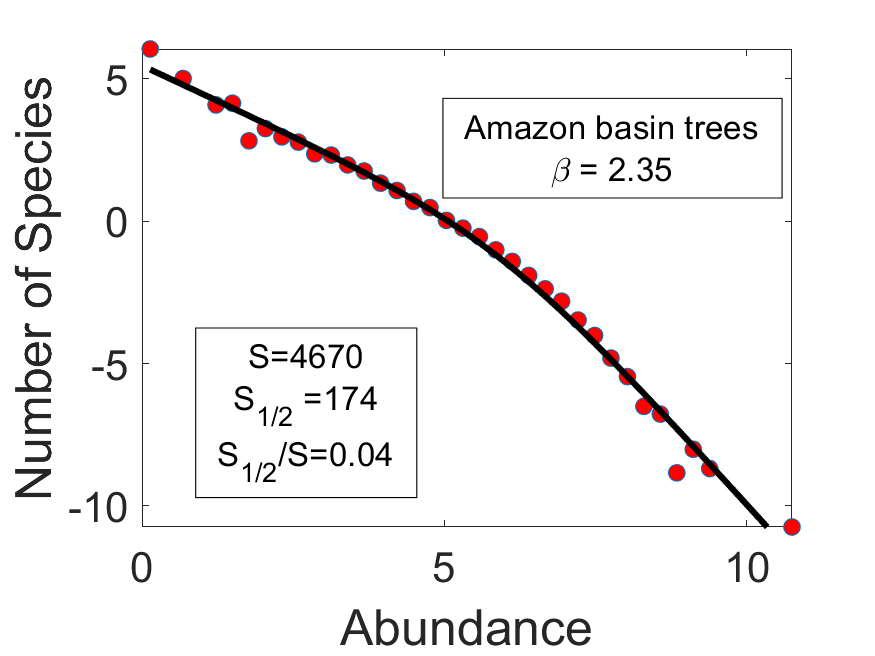} }
	\caption{Species abundance distribution, as obtained for a few diverse communities. The gut microbial community (OTUs of subject A from \citet{david2014host}) fits our formula \ref{eq5} quite well. Immigration is extremely weak, but the sampling power is strong enough to reveal some of the effects of the decrease in the number of species at low densities due to immigration.   The corresponding distribution for the oceanic prokaryote population~\cite{eguiluz2019scaling} is very close to a pure power law, possibly because the sampling is not deep enough. In both cases, $\beta < 2$, indicating that the community is dominated by only a few species. The distributions for the global bird population~\cite{callaghan2021global} and tropical trees in the Amazon Basin~\cite{cooper2024consistent} show a transition between two power-law behaviors. The overall number of bird and tree individuals is much smaller than that of microorganisms, so the effect of demographic stochasticity must be considered. When this is done (see Appendix \ref{SuppE}), we obtain an excellent fit for the results using the corrected formula (\ref{eq7}). For birds, we find $\beta \approx 2$, placing them on the margin between egalitarian and dominance communities. For tropical trees in the Amazon Basin, $\beta$ is definitely larger than two, indicating that the community is indeed egalitarian. Note the similarities between the values of $S_{1/2}/S$ in the empirical results and the numerical experiments in Fig. \ref{fig2}.   \label{fig3}}
\end{figure}

Figure \ref{fig3} shows that the formula we presented for a time-averaged neutral community with immigration and purely environmental stochasticity, Eq. (\ref{eq5}), describes the species abundance distribution for the gut microbiome quite well. In the case of ocean prokaryotes, only the power law in the tail is observed, perhaps because the sampling strength is insufficient~\cite{maruvka2010universal}.  On the other hand, for trees in tropical forest or birds, the distribution of population sizes shows a crossover between two descending power laws, without a maximum point at a finite abundance. 

Trees and birds differ from microorganisms in two important ways. First, population sizes (the number of individuals in each population) are astronomical in microorganisms and much smaller in macroorganisms. Second, since macroorganisms are physically larger, they are more resistant to environmental variations (they have mechanisms that buffer these variations). These two factors lead to a greater need to consider demographic stochasticity in macroorganisms -- both because local population sizes are not large and because the impact of environmental stochasticity on reproductive success is smaller.

In SI \ref{SuppE} we consider the case of a time-averaged neutral community with  demographic stochasticity, in addition to the previously considered immigration and environmental stochasticity. The expected distribution for species abundances now takes the form,
\begin{equation} \label{eq7}
P(n) = A n^{-1-2\mu/\sigma_d^2} \left(\frac{\sigma_d^2}{\sigma_e^2} + n\right)^{-2d/\sigma_e^2 - 2\mu/\sigma_d^2}.
\end{equation}
This expression converges to Eq. (\ref{eq5}) above in the limit $\sigma_d \ll \sigma_e$. This is a reasonable limit for microorganisms, but apparently not for macroorganisms. 

Still, the distinction between dominance and egalitarian communities depends solely on the decay of the tail and its corresponding exponent $\beta$, which, for Eq. (\ref{eq7}), remains as $\beta = -1-2d/\sigma_e^2$. Figure \ref{fig3} thus illustrates the three types of behavior demonstrated in the numerical experiments shown in Figure \ref{fig2}: The microorganism communities are in the dominance phase, the tropical tree community is in the egalitarian phase, and the bird community appears to fall in between.

A comparison between the values of the evenness parameter in Figure \ref{fig2}  above and the values of $S_{1/2}/S$ in \cite{van2024tiny} (for the same range of species richness $S$) also shows that some communities fall within the dominance phase, while others fall within the egalitarian phase. On the other hand, the results for plankton~\cite{ser2018ubiquitous}, in which $\beta \in [1..2]$, appear to suggest that these microorganismal communities are all in the dominance phase.

\section{Discussion and conclusions}

Hyperdiverse communities, like those analyzed in this paper, are extremely important and frequently occur in nature. However, quantifying their specific parameters is an impossible task. Therefore, the attempt to understand the factors that dictate the community structure in these systems requires the use of models, which should preferably be as generic as possible.

As mentioned in the introduction,  these models are divided into two classes: weak coupling models in which a small niche overlap is assumed, meaning that the interspecies interactions are much weaker than the intraspecies ones, and strong coupling models in which  niche overlap is assumed to be substantial. For example, in Generalized Lotka-Volterra models for $S \gg 1$ species,
\begin{equation} 
\frac{d N_i}{dt} = N_i \left( 1 - N_i - \sum_{j \neq i} \alpha_{i,j} N_j \right),
\end{equation}
the $\alpha_{i,j}$s are ${\cal O}(1)$ in the strong coupling scenario, and  ${\cal O}(1/S)$ in the weak coupling scenario.

 In the case of weakly interacting species, it is clear that the small impact of any given species leads to an effective single-species model that is a slight perturbation of the logistic dynamics, leading to a truncated Gaussian distribution.  Remarkably, even in the strongly interacting case, if the individual species abundances fluctuate strongly in time
(whether due to chaotic dynamics of the system~\cite{arnoulx2024many}, strong environmental stochasticity, or variability of the $\alpha_{i,j}$s over time~\cite{azaele2023large}), each species can be treated separately, with the sum of the small, uncorrelated effects of competition with other species represented by environmental stochasticity, characterized by typical magnitudes of strength and correlation time. This high niche overlap scenario yields a different type of effective single-species model, where the linear growth rate is weak and negative. This dynamics, which emphasizes the pronounced effect of stickiness, was considered here and in other recent works~\cite{danino2018theory, van2024tiny, mallmin2024chaotic}.

One might wonder, therefore, which of these two types of models is more suitable for describing highly diverse communities in nature. If one assumes (like Hutchinson in the paradox of the plankton~\cite{hutchinson1961paradox}) that the main competitive effect stems from the struggle for limiting resources such as water or food, it would be reasonable to believe that natural communities are closer to the realm of high niche overlap and strong couplings.

A separate question is the primary driver of the environmental stochasticity seen in the ecological data. An examination of time series for a single species in diverse ecological systems usually reveals environmental stochasticity and not much more than that~\cite{lande2003stochastic,kalyuzhny2014niche}. This phenomenon persists even in particularly careful laboratory experiments where environmental conditions are kept as constant as conceivably possible~\cite{hekstra2012contingency}.  Does it arise from chaotic behavior of the system (as suggested in~\cite{mallmin2024chaotic}), or from environmental fluctuations as we have assumed here? Alternatively, perhaps the source lies in the interplay between consumers and resources, as in~\cite{batista2021path}? 

It may be that to address this question, we will need to focus specifically on the dynamics of the most highly abundant species.  If the noise is intrinsic (chaos), stemming from the fluctuating influence of other species in the community, a species that has reached dominance would likely show completely different characteristics of abundance fluctuations compared to when that same species is at rarity. On the other hand, if the fluctuations are due to changes in weather or precipitation, we would expect abundance-independent statistics of abundance variations, as found empirically in~\cite{kalyuzhny2014niche}. In any case, while this question has some theoretical value, it is hard to believe it would have significant practical implications. After all, it is well known that distinguishing between high-dimensional chaos and mere noise is virtually impossible.

 Whatever the underlying mechanism, the net result, as we have demonstrated herein, is that in the limit of  strong interspecies coupling, a transition occurs between an egalitarian phase and a dominance phase. When the immigration rate $\mu$  is reduced, total carrying capacity $K$ increases,  or when environmental stochasticity $\sigma_e$ increases, the system can suddenly lose a significant amount of diversity at the transition point. The dependence on the total population size $K$ is particularly interesting, and has to do with the dispute about the relationships between productivity and species richness~\cite{waide1999relationship,kadmon2006effects}.

We will now discuss some potential extensions of the model described here and explore their possible implications.

Our model assumed an uncorrelated response of species to environmental variations. This treatment can be easily extended to include {\it correlated responses}, using techniques similar to those implemented by ~\citet{loreau2008species}. In general, under correlated responses, the effective number of species in the community decreases.

Another interesting point relates to the interplay between the stickiness mechanism, through which environmental stochasticity causes populations to spend long periods in a state of rarity, and mechanisms such as the { \it storage effect or relative nonlinearity}~\cite{chesson1982storage,chesson2000mechanisms,ellner2016quantify,letten2018species} that allow rare populations to invade due to environmental stochasticity. It is likely that these mechanisms weaken as the number of species increases (for storage, this has been demonstrated in several studies~\cite{chesson1989short,pande2022temporal}), and therefore, at least in diverse communities of competing species, the dominant effect will actually be that of stickiness.

The time-averaged neutral dynamic considered here assumes that the differences in average fitness between species are negligible, at least to a first approximation. This assumption is necessary in situations where niche overlap is large; otherwise, fitness differences would cause the extinction of most species. The justification for this can come from processes leading to emergent neutrality~\cite{holt2006emergent,vergnon2012emergent}, or from the fact that environmental stochasticity itself is also a mechanism that "neutralizes" fitness differences~\cite{pande2022temporal}. Extending out model to include {\it weak deviations} from time-averaged neutrality, perhaps using the dynamic mean-field approximation~\cite{roy2019numerical} is a tempting possibility. 

Species coexistence has long been, and remains, a theoretical puzzle of immense importance for understanding the dynamics of biological systems, with far-reaching practical implications. The research conducted in recent years has provided powerful theoretical tools that allow us to focus the discussion and understand the generic implications of community structure and the nature of species interactions on the range of possible outcomes. We believe that the parameter range we have addressed in this paper — strong coupling and strong stochasticity — is relevant for a wide variety of ecological systems, and we hope that our work will serve as a foundation for further studies that explain the wide range of diversity levels (for example, between a tropical forest and a tundra, or between different types of microbiome) in relation to the phase transition described here.

{\bf Acknowledgments:} We thank Emil Mallmin and Egbert van Nes for the interesting exchange of ideas that contributed to the shaping of this work.

\bibliography{ref}

\begin{thebibliography}{60}%
\makeatletter
\providecommand \@ifxundefined [1]{%
 \@ifx{#1\undefined}
}%
\providecommand \@ifnum [1]{%
 \ifnum #1\expandafter \@firstoftwo
 \else \expandafter \@secondoftwo
 \fi
}%
\providecommand \@ifx [1]{%
 \ifx #1\expandafter \@firstoftwo
 \else \expandafter \@secondoftwo
 \fi
}%
\providecommand \natexlab [1]{#1}%
\providecommand \enquote  [1]{``#1''}%
\providecommand \bibnamefont  [1]{#1}%
\providecommand \bibfnamefont [1]{#1}%
\providecommand \citenamefont [1]{#1}%
\providecommand \href@noop [0]{\@secondoftwo}%
\providecommand \href [0]{\begingroup \@sanitize@url \@href}%
\providecommand \@href[1]{\@@startlink{#1}\@@href}%
\providecommand \@@href[1]{\endgroup#1\@@endlink}%
\providecommand \@sanitize@url [0]{\catcode `\\12\catcode `\$12\catcode `\&12\catcode `\#12\catcode `\^12\catcode `\_12\catcode `\%12\relax}%
\providecommand \@@startlink[1]{}%
\providecommand \@@endlink[0]{}%
\providecommand \url  [0]{\begingroup\@sanitize@url \@url }%
\providecommand \@url [1]{\endgroup\@href {#1}{\urlprefix }}%
\providecommand \urlprefix  [0]{URL }%
\providecommand \Eprint [0]{\href }%
\providecommand \doibase [0]{https://doi.org/}%
\providecommand \selectlanguage [0]{\@gobble}%
\providecommand \bibinfo  [0]{\@secondoftwo}%
\providecommand \bibfield  [0]{\@secondoftwo}%
\providecommand \translation [1]{[#1]}%
\providecommand \BibitemOpen [0]{}%
\providecommand \bibitemStop [0]{}%
\providecommand \bibitemNoStop [0]{.\EOS\space}%
\providecommand \EOS [0]{\spacefactor3000\relax}%
\providecommand \BibitemShut  [1]{\csname bibitem#1\endcsname}%
\let\auto@bib@innerbib\@empty
\bibitem [{\citenamefont {Hutchinson}(1961{\natexlab{a}})}]{plankton}%
  \BibitemOpen
  \bibfield  {author} {\bibinfo {author} {\bibfnamefont {G.~E.}\ \bibnamefont {Hutchinson}},\ }\bibfield  {title} {\bibinfo {title} {The paradox of the plankton},\ }\href@noop {} {\bibfield  {journal} {\bibinfo  {journal} {The American Naturalist}\ }\textbf {\bibinfo {volume} {95}},\ \bibinfo {pages} {137} (\bibinfo {year} {1961}{\natexlab{a}})}\BibitemShut {NoStop}%
\bibitem [{\citenamefont {Stomp}\ \emph {et~al.}(2011)\citenamefont {Stomp}, \citenamefont {Huisman}, \citenamefont {Mittelbach}, \citenamefont {Litchman},\ and\ \citenamefont {Klausmeier}}]{stomp2011large}%
  \BibitemOpen
  \bibfield  {author} {\bibinfo {author} {\bibfnamefont {M.}~\bibnamefont {Stomp}}, \bibinfo {author} {\bibfnamefont {J.}~\bibnamefont {Huisman}}, \bibinfo {author} {\bibfnamefont {G.~G.}\ \bibnamefont {Mittelbach}}, \bibinfo {author} {\bibfnamefont {E.}~\bibnamefont {Litchman}},\ and\ \bibinfo {author} {\bibfnamefont {C.~A.}\ \bibnamefont {Klausmeier}},\ }\bibfield  {title} {\bibinfo {title} {Large-scale biodiversity patterns in freshwater phytoplankton},\ }\href@noop {} {\bibfield  {journal} {\bibinfo  {journal} {Ecology}\ }\textbf {\bibinfo {volume} {92}},\ \bibinfo {pages} {2096} (\bibinfo {year} {2011})}\BibitemShut {NoStop}%
\bibitem [{\citenamefont {Ter~Steege}\ \emph {et~al.}(2013)\citenamefont {Ter~Steege}, \citenamefont {Pitman}, \citenamefont {Sabatier}, \citenamefont {Baraloto}, \citenamefont {Salom{\~a}o}, \citenamefont {Guevara}, \citenamefont {Phillips}, \citenamefont {Castilho}, \citenamefont {Magnusson}, \citenamefont {Molino} \emph {et~al.}}]{ter2013hyperdominance}%
  \BibitemOpen
  \bibfield  {author} {\bibinfo {author} {\bibfnamefont {H.}~\bibnamefont {Ter~Steege}}, \bibinfo {author} {\bibfnamefont {N.~C.}\ \bibnamefont {Pitman}}, \bibinfo {author} {\bibfnamefont {D.}~\bibnamefont {Sabatier}}, \bibinfo {author} {\bibfnamefont {C.}~\bibnamefont {Baraloto}}, \bibinfo {author} {\bibfnamefont {R.~P.}\ \bibnamefont {Salom{\~a}o}}, \bibinfo {author} {\bibfnamefont {J.~E.}\ \bibnamefont {Guevara}}, \bibinfo {author} {\bibfnamefont {O.~L.}\ \bibnamefont {Phillips}}, \bibinfo {author} {\bibfnamefont {C.~V.}\ \bibnamefont {Castilho}}, \bibinfo {author} {\bibfnamefont {W.~E.}\ \bibnamefont {Magnusson}}, \bibinfo {author} {\bibfnamefont {J.-F.}\ \bibnamefont {Molino}}, \emph {et~al.},\ }\bibfield  {title} {\bibinfo {title} {Hyperdominance in the amazonian tree flora},\ }\href@noop {} {\bibfield  {journal} {\bibinfo  {journal} {Science}\ }\textbf {\bibinfo {volume} {342}},\ \bibinfo {pages} {1243092} (\bibinfo {year} {2013})}\BibitemShut {NoStop}%
\bibitem [{\citenamefont {Connolly}\ \emph {et~al.}(2014)\citenamefont {Connolly}, \citenamefont {MacNeil}, \citenamefont {Caley}, \citenamefont {Knowlton}, \citenamefont {Cripps}, \citenamefont {Hisano}, \citenamefont {Thibaut}, \citenamefont {Bhattacharya}, \citenamefont {Benedetti-Cecchi}, \citenamefont {Brainard} \emph {et~al.}}]{connolly2014commonness}%
  \BibitemOpen
  \bibfield  {author} {\bibinfo {author} {\bibfnamefont {S.~R.}\ \bibnamefont {Connolly}}, \bibinfo {author} {\bibfnamefont {M.~A.}\ \bibnamefont {MacNeil}}, \bibinfo {author} {\bibfnamefont {M.~J.}\ \bibnamefont {Caley}}, \bibinfo {author} {\bibfnamefont {N.}~\bibnamefont {Knowlton}}, \bibinfo {author} {\bibfnamefont {E.}~\bibnamefont {Cripps}}, \bibinfo {author} {\bibfnamefont {M.}~\bibnamefont {Hisano}}, \bibinfo {author} {\bibfnamefont {L.~M.}\ \bibnamefont {Thibaut}}, \bibinfo {author} {\bibfnamefont {B.~D.}\ \bibnamefont {Bhattacharya}}, \bibinfo {author} {\bibfnamefont {L.}~\bibnamefont {Benedetti-Cecchi}}, \bibinfo {author} {\bibfnamefont {R.~E.}\ \bibnamefont {Brainard}}, \emph {et~al.},\ }\bibfield  {title} {\bibinfo {title} {Commonness and rarity in the marine biosphere},\ }\href@noop {} {\bibfield  {journal} {\bibinfo  {journal} {Proceedings of the National Academy of Sciences}\ }\textbf {\bibinfo {volume} {111}},\ \bibinfo {pages} {8524} (\bibinfo {year} {2014})}\BibitemShut {NoStop}%
\bibitem [{\citenamefont {Fierer}\ \emph {et~al.}(2007)\citenamefont {Fierer}, \citenamefont {Breitbart}, \citenamefont {Nulton}, \citenamefont {Salamon}, \citenamefont {Lozupone}, \citenamefont {Jones}, \citenamefont {Robeson}, \citenamefont {Edwards}, \citenamefont {Felts}, \citenamefont {Rayhawk} \emph {et~al.}}]{fierer2007metagenomic}%
  \BibitemOpen
  \bibfield  {author} {\bibinfo {author} {\bibfnamefont {N.}~\bibnamefont {Fierer}}, \bibinfo {author} {\bibfnamefont {M.}~\bibnamefont {Breitbart}}, \bibinfo {author} {\bibfnamefont {J.}~\bibnamefont {Nulton}}, \bibinfo {author} {\bibfnamefont {P.}~\bibnamefont {Salamon}}, \bibinfo {author} {\bibfnamefont {C.}~\bibnamefont {Lozupone}}, \bibinfo {author} {\bibfnamefont {R.}~\bibnamefont {Jones}}, \bibinfo {author} {\bibfnamefont {M.}~\bibnamefont {Robeson}}, \bibinfo {author} {\bibfnamefont {R.~A.}\ \bibnamefont {Edwards}}, \bibinfo {author} {\bibfnamefont {B.}~\bibnamefont {Felts}}, \bibinfo {author} {\bibfnamefont {S.}~\bibnamefont {Rayhawk}}, \emph {et~al.},\ }\bibfield  {title} {\bibinfo {title} {Metagenomic and small-subunit rrna analyses reveal the genetic diversity of bacteria, archaea, fungi, and viruses in soil},\ }\href@noop {} {\bibfield  {journal} {\bibinfo  {journal} {Applied and environmental microbiology}\ }\textbf {\bibinfo {volume} {73}},\ \bibinfo {pages} {7059} (\bibinfo {year}
  {2007})}\BibitemShut {NoStop}%
\bibitem [{\citenamefont {Gause}(2003)}]{gause2003struggle}%
  \BibitemOpen
  \bibfield  {author} {\bibinfo {author} {\bibfnamefont {G.~F.}\ \bibnamefont {Gause}},\ }\href@noop {} {\emph {\bibinfo {title} {The Struggle for Existence}}}\ (\bibinfo  {publisher} {Courier Corporation},\ \bibinfo {year} {2003})\BibitemShut {NoStop}%
\bibitem [{\citenamefont {Tilman}(1982)}]{tilman1982resource}%
  \BibitemOpen
  \bibfield  {author} {\bibinfo {author} {\bibfnamefont {D.}~\bibnamefont {Tilman}},\ }\href@noop {} {\emph {\bibinfo {title} {Resource Competition and Community Structure}}},\ \bibinfo {number} {17}\ (\bibinfo  {publisher} {Princeton university press},\ \bibinfo {year} {1982})\BibitemShut {NoStop}%
\bibitem [{\citenamefont {Chesson}(2000)}]{chesson2000mechanisms}%
  \BibitemOpen
  \bibfield  {author} {\bibinfo {author} {\bibfnamefont {P.~L.}\ \bibnamefont {Chesson}},\ }\bibfield  {title} {\bibinfo {title} {Mechanisms of maintenance of species diversity},\ }\href@noop {} {\bibfield  {journal} {\bibinfo  {journal} {Annual Review of Ecology and Systematics}\ }\textbf {\bibinfo {volume} {31}},\ \bibinfo {pages} {343} (\bibinfo {year} {2000})}\BibitemShut {NoStop}%
\bibitem [{\citenamefont {Chesson}(2003)}]{chesson2003quantifying}%
  \BibitemOpen
  \bibfield  {author} {\bibinfo {author} {\bibfnamefont {P.~L.}\ \bibnamefont {Chesson}},\ }\bibfield  {title} {\bibinfo {title} {Quantifying and testing coexistence mechanisms arising from recruitment fluctuations},\ }\href@noop {} {\bibfield  {journal} {\bibinfo  {journal} {Theoretical Population Biology}\ }\textbf {\bibinfo {volume} {64}},\ \bibinfo {pages} {345} (\bibinfo {year} {2003})}\BibitemShut {NoStop}%
\bibitem [{\citenamefont {David}\ \emph {et~al.}(2014)\citenamefont {David}, \citenamefont {Materna}, \citenamefont {Friedman}, \citenamefont {Campos-Baptista}, \citenamefont {Blackburn}, \citenamefont {Perrotta}, \citenamefont {Erdman},\ and\ \citenamefont {Alm}}]{david2014host}%
  \BibitemOpen
  \bibfield  {author} {\bibinfo {author} {\bibfnamefont {L.~A.}\ \bibnamefont {David}}, \bibinfo {author} {\bibfnamefont {A.~C.}\ \bibnamefont {Materna}}, \bibinfo {author} {\bibfnamefont {J.}~\bibnamefont {Friedman}}, \bibinfo {author} {\bibfnamefont {M.~I.}\ \bibnamefont {Campos-Baptista}}, \bibinfo {author} {\bibfnamefont {M.~C.}\ \bibnamefont {Blackburn}}, \bibinfo {author} {\bibfnamefont {A.}~\bibnamefont {Perrotta}}, \bibinfo {author} {\bibfnamefont {S.~E.}\ \bibnamefont {Erdman}},\ and\ \bibinfo {author} {\bibfnamefont {E.~J.}\ \bibnamefont {Alm}},\ }\bibfield  {title} {\bibinfo {title} {Host lifestyle affects human microbiota on daily timescales},\ }\href@noop {} {\bibfield  {journal} {\bibinfo  {journal} {Genome biology}\ }\textbf {\bibinfo {volume} {15}},\ \bibinfo {pages} {1} (\bibinfo {year} {2014})}\BibitemShut {NoStop}%
\bibitem [{\citenamefont {Grilli}(2020)}]{grilli2020macroecological}%
  \BibitemOpen
  \bibfield  {author} {\bibinfo {author} {\bibfnamefont {J.}~\bibnamefont {Grilli}},\ }\bibfield  {title} {\bibinfo {title} {Macroecological laws describe variation and diversity in microbial communities},\ }\href@noop {} {\bibfield  {journal} {\bibinfo  {journal} {Nature communications}\ }\textbf {\bibinfo {volume} {11}},\ \bibinfo {pages} {4743} (\bibinfo {year} {2020})}\BibitemShut {NoStop}%
\bibitem [{\citenamefont {Egu{\'\i}luz}\ \emph {et~al.}(2019)\citenamefont {Egu{\'\i}luz}, \citenamefont {Salazar}, \citenamefont {Fern{\'a}ndez-Gracia}, \citenamefont {Pearman}, \citenamefont {Gasol}, \citenamefont {Acinas}, \citenamefont {Sunagawa}, \citenamefont {Irigoien},\ and\ \citenamefont {Duarte}}]{eguiluz2019scaling}%
  \BibitemOpen
  \bibfield  {author} {\bibinfo {author} {\bibfnamefont {V.~M.}\ \bibnamefont {Egu{\'\i}luz}}, \bibinfo {author} {\bibfnamefont {G.}~\bibnamefont {Salazar}}, \bibinfo {author} {\bibfnamefont {J.}~\bibnamefont {Fern{\'a}ndez-Gracia}}, \bibinfo {author} {\bibfnamefont {J.~K.}\ \bibnamefont {Pearman}}, \bibinfo {author} {\bibfnamefont {J.~M.}\ \bibnamefont {Gasol}}, \bibinfo {author} {\bibfnamefont {S.~G.}\ \bibnamefont {Acinas}}, \bibinfo {author} {\bibfnamefont {S.}~\bibnamefont {Sunagawa}}, \bibinfo {author} {\bibfnamefont {X.}~\bibnamefont {Irigoien}},\ and\ \bibinfo {author} {\bibfnamefont {C.~M.}\ \bibnamefont {Duarte}},\ }\bibfield  {title} {\bibinfo {title} {Scaling of species distribution explains the vast potential marine prokaryote diversity},\ }\href@noop {} {\bibfield  {journal} {\bibinfo  {journal} {Scientific reports}\ }\textbf {\bibinfo {volume} {9}},\ \bibinfo {pages} {18710} (\bibinfo {year} {2019})}\BibitemShut {NoStop}%
\bibitem [{\citenamefont {Cooper}\ \emph {et~al.}(2024)\citenamefont {Cooper}, \citenamefont {Lewis}, \citenamefont {Sullivan}, \citenamefont {Prado}, \citenamefont {Ter~Steege}, \citenamefont {Barbier}, \citenamefont {Slik}, \citenamefont {Sonk{\'e}}, \citenamefont {Ewango}, \citenamefont {Adu-Bredu} \emph {et~al.}}]{cooper2024consistent}%
  \BibitemOpen
  \bibfield  {author} {\bibinfo {author} {\bibfnamefont {D.~L.}\ \bibnamefont {Cooper}}, \bibinfo {author} {\bibfnamefont {S.~L.}\ \bibnamefont {Lewis}}, \bibinfo {author} {\bibfnamefont {M.~J.}\ \bibnamefont {Sullivan}}, \bibinfo {author} {\bibfnamefont {P.~I.}\ \bibnamefont {Prado}}, \bibinfo {author} {\bibfnamefont {H.}~\bibnamefont {Ter~Steege}}, \bibinfo {author} {\bibfnamefont {N.}~\bibnamefont {Barbier}}, \bibinfo {author} {\bibfnamefont {F.}~\bibnamefont {Slik}}, \bibinfo {author} {\bibfnamefont {B.}~\bibnamefont {Sonk{\'e}}}, \bibinfo {author} {\bibfnamefont {C.~E.}\ \bibnamefont {Ewango}}, \bibinfo {author} {\bibfnamefont {S.}~\bibnamefont {Adu-Bredu}}, \emph {et~al.},\ }\bibfield  {title} {\bibinfo {title} {Consistent patterns of common species across tropical tree communities},\ }\href@noop {} {\bibfield  {journal} {\bibinfo  {journal} {Nature}\ }\textbf {\bibinfo {volume} {625}},\ \bibinfo {pages} {728} (\bibinfo {year} {2024})}\BibitemShut {NoStop}%
\bibitem [{\citenamefont {Callaghan}\ \emph {et~al.}(2021)\citenamefont {Callaghan}, \citenamefont {Nakagawa},\ and\ \citenamefont {Cornwell}}]{callaghan2021global}%
  \BibitemOpen
  \bibfield  {author} {\bibinfo {author} {\bibfnamefont {C.~T.}\ \bibnamefont {Callaghan}}, \bibinfo {author} {\bibfnamefont {S.}~\bibnamefont {Nakagawa}},\ and\ \bibinfo {author} {\bibfnamefont {W.~K.}\ \bibnamefont {Cornwell}},\ }\bibfield  {title} {\bibinfo {title} {Global abundance estimates for 9,700 bird species},\ }\href@noop {} {\bibfield  {journal} {\bibinfo  {journal} {Proceedings of the National Academy of Sciences}\ }\textbf {\bibinfo {volume} {118}},\ \bibinfo {pages} {e2023170118} (\bibinfo {year} {2021})}\BibitemShut {NoStop}%
\bibitem [{\citenamefont {May}(1972)}]{may1972will}%
  \BibitemOpen
  \bibfield  {author} {\bibinfo {author} {\bibfnamefont {R.~M.}\ \bibnamefont {May}},\ }\bibfield  {title} {\bibinfo {title} {Will a large complex system be stable?},\ }\href@noop {} {\bibfield  {journal} {\bibinfo  {journal} {Nature}\ }\textbf {\bibinfo {volume} {238}},\ \bibinfo {pages} {413} (\bibinfo {year} {1972})}\BibitemShut {NoStop}%
\bibitem [{\citenamefont {Fisher}\ and\ \citenamefont {Mehta}(2014)}]{fisher2014transition}%
  \BibitemOpen
  \bibfield  {author} {\bibinfo {author} {\bibfnamefont {C.~K.}\ \bibnamefont {Fisher}}\ and\ \bibinfo {author} {\bibfnamefont {P.}~\bibnamefont {Mehta}},\ }\bibfield  {title} {\bibinfo {title} {The transition between the niche and neutral regimes in ecology},\ }\href@noop {} {\bibfield  {journal} {\bibinfo  {journal} {Proceedings of the National Academy of Sciences}\ }\textbf {\bibinfo {volume} {111}},\ \bibinfo {pages} {13111} (\bibinfo {year} {2014})}\BibitemShut {NoStop}%
\bibitem [{\citenamefont {Kessler}\ and\ \citenamefont {Shnerb}(2015)}]{kessler2015generalized}%
  \BibitemOpen
  \bibfield  {author} {\bibinfo {author} {\bibfnamefont {D.~A.}\ \bibnamefont {Kessler}}\ and\ \bibinfo {author} {\bibfnamefont {N.~M.}\ \bibnamefont {Shnerb}},\ }\bibfield  {title} {\bibinfo {title} {Generalized model of island biodiversity},\ }\href@noop {} {\bibfield  {journal} {\bibinfo  {journal} {Physical Review E}\ }\textbf {\bibinfo {volume} {91}},\ \bibinfo {pages} {042705} (\bibinfo {year} {2015})}\BibitemShut {NoStop}%
\bibitem [{\citenamefont {Bunin}(2017)}]{bunin2017ecological}%
  \BibitemOpen
  \bibfield  {author} {\bibinfo {author} {\bibfnamefont {G.}~\bibnamefont {Bunin}},\ }\bibfield  {title} {\bibinfo {title} {Ecological communities with lotka-volterra dynamics},\ }\href@noop {} {\bibfield  {journal} {\bibinfo  {journal} {Physical Review E}\ }\textbf {\bibinfo {volume} {95}},\ \bibinfo {pages} {042414} (\bibinfo {year} {2017})}\BibitemShut {NoStop}%
\bibitem [{\citenamefont {Barbier}\ \emph {et~al.}(2018)\citenamefont {Barbier}, \citenamefont {Arnoldi}, \citenamefont {Bunin},\ and\ \citenamefont {Loreau}}]{barbier2018generic}%
  \BibitemOpen
  \bibfield  {author} {\bibinfo {author} {\bibfnamefont {M.}~\bibnamefont {Barbier}}, \bibinfo {author} {\bibfnamefont {J.-F.}\ \bibnamefont {Arnoldi}}, \bibinfo {author} {\bibfnamefont {G.}~\bibnamefont {Bunin}},\ and\ \bibinfo {author} {\bibfnamefont {M.}~\bibnamefont {Loreau}},\ }\bibfield  {title} {\bibinfo {title} {Generic assembly patterns in complex ecological communities},\ }\href@noop {} {\bibfield  {journal} {\bibinfo  {journal} {Proceedings of the National Academy of Sciences}\ }\textbf {\bibinfo {volume} {115}},\ \bibinfo {pages} {2156} (\bibinfo {year} {2018})}\BibitemShut {NoStop}%
\bibitem [{\citenamefont {van Nes}\ \emph {et~al.}(2024)\citenamefont {van Nes}, \citenamefont {Pujoni}, \citenamefont {Shetty}, \citenamefont {Straatsma}, \citenamefont {de~Vos},\ and\ \citenamefont {Scheffer}}]{van2024tiny}%
  \BibitemOpen
  \bibfield  {author} {\bibinfo {author} {\bibfnamefont {E.~H.}\ \bibnamefont {van Nes}}, \bibinfo {author} {\bibfnamefont {D.~G.}\ \bibnamefont {Pujoni}}, \bibinfo {author} {\bibfnamefont {S.~A.}\ \bibnamefont {Shetty}}, \bibinfo {author} {\bibfnamefont {G.}~\bibnamefont {Straatsma}}, \bibinfo {author} {\bibfnamefont {W.~M.}\ \bibnamefont {de~Vos}},\ and\ \bibinfo {author} {\bibfnamefont {M.}~\bibnamefont {Scheffer}},\ }\bibfield  {title} {\bibinfo {title} {A tiny fraction of all species forms most of nature: Rarity as a sticky state},\ }\href@noop {} {\bibfield  {journal} {\bibinfo  {journal} {Proceedings of the National Academy of Sciences}\ }\textbf {\bibinfo {volume} {121}},\ \bibinfo {pages} {e2221791120} (\bibinfo {year} {2024})}\BibitemShut {NoStop}%
\bibitem [{\citenamefont {Azaele}\ and\ \citenamefont {Maritan}(2023)}]{azaele2023large}%
  \BibitemOpen
  \bibfield  {author} {\bibinfo {author} {\bibfnamefont {S.}~\bibnamefont {Azaele}}\ and\ \bibinfo {author} {\bibfnamefont {A.}~\bibnamefont {Maritan}},\ }\bibfield  {title} {\bibinfo {title} {Large system population dynamics with non-gaussian interactions},\ }\href@noop {} {\bibfield  {journal} {\bibinfo  {journal} {arXiv preprint arXiv:2306.13449}\ } (\bibinfo {year} {2023})}\BibitemShut {NoStop}%
\bibitem [{\citenamefont {Marcus}\ \emph {et~al.}(2022)\citenamefont {Marcus}, \citenamefont {Turner},\ and\ \citenamefont {Bunin}}]{marcus2022local}%
  \BibitemOpen
  \bibfield  {author} {\bibinfo {author} {\bibfnamefont {S.}~\bibnamefont {Marcus}}, \bibinfo {author} {\bibfnamefont {A.~M.}\ \bibnamefont {Turner}},\ and\ \bibinfo {author} {\bibfnamefont {G.}~\bibnamefont {Bunin}},\ }\bibfield  {title} {\bibinfo {title} {Local and collective transitions in sparsely-interacting ecological communities},\ }\href@noop {} {\bibfield  {journal} {\bibinfo  {journal} {PLoS computational biology}\ }\textbf {\bibinfo {volume} {18}},\ \bibinfo {pages} {e1010274} (\bibinfo {year} {2022})}\BibitemShut {NoStop}%
\bibitem [{\citenamefont {Crow}\ \emph {et~al.}(1970)\citenamefont {Crow}, \citenamefont {Kimura} \emph {et~al.}}]{crow1970introduction}%
  \BibitemOpen
  \bibfield  {author} {\bibinfo {author} {\bibfnamefont {J.~F.}\ \bibnamefont {Crow}}, \bibinfo {author} {\bibfnamefont {M.}~\bibnamefont {Kimura}}, \emph {et~al.},\ }\href@noop {} {\emph {\bibinfo {title} {An Introduction to Population Genetics Theory}}}\ (\bibinfo  {publisher} {New York, Evanston and London: Harper \& Row, Publishers},\ \bibinfo {year} {1970})\BibitemShut {NoStop}%
\bibitem [{\citenamefont {Hubbell}(2001)}]{hubbell_book}%
  \BibitemOpen
  \bibfield  {author} {\bibinfo {author} {\bibfnamefont {S.~P.}\ \bibnamefont {Hubbell}},\ }\href@noop {} {\emph {\bibinfo {title} {The Unified Neutral Theory of Biodiversity and Biogeography}}},\ Monographs in Population Biology 32\ (\bibinfo  {publisher} {Princeton University Press},\ \bibinfo {address} {Princeton, N.J.},\ \bibinfo {year} {2001})\BibitemShut {NoStop}%
\bibitem [{\citenamefont {Volkov}\ \emph {et~al.}(2003)\citenamefont {Volkov}, \citenamefont {Banavar}, \citenamefont {Hubbell},\ and\ \citenamefont {Maritan}}]{maritan1}%
  \BibitemOpen
  \bibfield  {author} {\bibinfo {author} {\bibfnamefont {I.}~\bibnamefont {Volkov}}, \bibinfo {author} {\bibfnamefont {J.~R.}\ \bibnamefont {Banavar}}, \bibinfo {author} {\bibfnamefont {S.~P.}\ \bibnamefont {Hubbell}},\ and\ \bibinfo {author} {\bibfnamefont {A.}~\bibnamefont {Maritan}},\ }\bibfield  {title} {\bibinfo {title} {Neutral theory and relative species abundance in ecology.},\ }\href@noop {} {\bibfield  {journal} {\bibinfo  {journal} {Nature}\ }\textbf {\bibinfo {volume} {424}},\ \bibinfo {pages} {1035} (\bibinfo {year} {2003})}\BibitemShut {NoStop}%
\bibitem [{\citenamefont {Azaele}\ \emph {et~al.}(2015)\citenamefont {Azaele}, \citenamefont {Maritan}, \citenamefont {Cornell}, \citenamefont {Suweis}, \citenamefont {Banavar}, \citenamefont {Gabriel},\ and\ \citenamefont {Kunin}}]{azaele2015towards}%
  \BibitemOpen
  \bibfield  {author} {\bibinfo {author} {\bibfnamefont {S.}~\bibnamefont {Azaele}}, \bibinfo {author} {\bibfnamefont {A.}~\bibnamefont {Maritan}}, \bibinfo {author} {\bibfnamefont {S.~J.}\ \bibnamefont {Cornell}}, \bibinfo {author} {\bibfnamefont {S.}~\bibnamefont {Suweis}}, \bibinfo {author} {\bibfnamefont {J.~R.}\ \bibnamefont {Banavar}}, \bibinfo {author} {\bibfnamefont {D.}~\bibnamefont {Gabriel}},\ and\ \bibinfo {author} {\bibfnamefont {W.~E.}\ \bibnamefont {Kunin}},\ }\bibfield  {title} {\bibinfo {title} {Towards a unified descriptive theory for spatial ecology: predicting biodiversity patterns across spatial scales},\ }\href@noop {} {\bibfield  {journal} {\bibinfo  {journal} {Methods in Ecology and Evolution}\ }\textbf {\bibinfo {volume} {6}},\ \bibinfo {pages} {324} (\bibinfo {year} {2015})}\BibitemShut {NoStop}%
\bibitem [{\citenamefont {Rosindell}\ \emph {et~al.}(2011)\citenamefont {Rosindell}, \citenamefont {Hubbell},\ and\ \citenamefont {Etienne}}]{TREE2011}%
  \BibitemOpen
  \bibfield  {author} {\bibinfo {author} {\bibfnamefont {J.}~\bibnamefont {Rosindell}}, \bibinfo {author} {\bibfnamefont {S.~P.}\ \bibnamefont {Hubbell}},\ and\ \bibinfo {author} {\bibfnamefont {R.~S.}\ \bibnamefont {Etienne}},\ }\bibfield  {title} {\bibinfo {title} {The unified neutral theory of biodiversity and biogeography at age ten},\ }\href@noop {} {\bibfield  {journal} {\bibinfo  {journal} {Trends in Ecology \& Evolution}\ }\textbf {\bibinfo {volume} {26}},\ \bibinfo {pages} {340} (\bibinfo {year} {2011})}\BibitemShut {NoStop}%
\bibitem [{\citenamefont {Nee}(2005)}]{nee2005neutral}%
  \BibitemOpen
  \bibfield  {author} {\bibinfo {author} {\bibfnamefont {S.}~\bibnamefont {Nee}},\ }\bibfield  {title} {\bibinfo {title} {The neutral theory of biodiversity: do the numbers add up?},\ }\href@noop {} {\bibfield  {journal} {\bibinfo  {journal} {Functional Ecology}\ }\textbf {\bibinfo {volume} {19}},\ \bibinfo {pages} {173} (\bibinfo {year} {2005})}\BibitemShut {NoStop}%
\bibitem [{\citenamefont {Ricklefs}(2006)}]{ricklefs2006unified}%
  \BibitemOpen
  \bibfield  {author} {\bibinfo {author} {\bibfnamefont {R.~E.}\ \bibnamefont {Ricklefs}},\ }\bibfield  {title} {\bibinfo {title} {The unified neutral theory of biodiversity: do the numbers add up?},\ }\href@noop {} {\bibfield  {journal} {\bibinfo  {journal} {Ecology}\ }\textbf {\bibinfo {volume} {87}},\ \bibinfo {pages} {1424} (\bibinfo {year} {2006})}\BibitemShut {NoStop}%
\bibitem [{\citenamefont {Leigh}(2007)}]{leigh2007neutral}%
  \BibitemOpen
  \bibfield  {author} {\bibinfo {author} {\bibfnamefont {E.~G.}\ \bibnamefont {Leigh}},\ }\bibfield  {title} {\bibinfo {title} {Neutral theory: a historical perspective},\ }\href@noop {} {\bibfield  {journal} {\bibinfo  {journal} {Journal of Evolutionary Biology}\ }\textbf {\bibinfo {volume} {20}},\ \bibinfo {pages} {2075} (\bibinfo {year} {2007})}\BibitemShut {NoStop}%
\bibitem [{\citenamefont {Chisholm}\ and\ \citenamefont {O'Dwyer}(2014)}]{chisholm2014species}%
  \BibitemOpen
  \bibfield  {author} {\bibinfo {author} {\bibfnamefont {R.~A.}\ \bibnamefont {Chisholm}}\ and\ \bibinfo {author} {\bibfnamefont {J.~P.}\ \bibnamefont {O'Dwyer}},\ }\bibfield  {title} {\bibinfo {title} {Species ages in neutral biodiversity models},\ }\href@noop {} {\bibfield  {journal} {\bibinfo  {journal} {Theoretical Population Biology}\ }\textbf {\bibinfo {volume} {93}},\ \bibinfo {pages} {85} (\bibinfo {year} {2014})}\BibitemShut {NoStop}%
\bibitem [{\citenamefont {Kalyuzhny}\ \emph {et~al.}(2014{\natexlab{a}})\citenamefont {Kalyuzhny}, \citenamefont {Seri}, \citenamefont {Chocron}, \citenamefont {Flather}, \citenamefont {Kadmon},\ and\ \citenamefont {Shnerb}}]{kalyuzhny2014niche}%
  \BibitemOpen
  \bibfield  {author} {\bibinfo {author} {\bibfnamefont {M.}~\bibnamefont {Kalyuzhny}}, \bibinfo {author} {\bibfnamefont {E.}~\bibnamefont {Seri}}, \bibinfo {author} {\bibfnamefont {R.}~\bibnamefont {Chocron}}, \bibinfo {author} {\bibfnamefont {C.~H.}\ \bibnamefont {Flather}}, \bibinfo {author} {\bibfnamefont {R.}~\bibnamefont {Kadmon}},\ and\ \bibinfo {author} {\bibfnamefont {N.~M.}\ \bibnamefont {Shnerb}},\ }\bibfield  {title} {\bibinfo {title} {Niche versus neutrality: a dynamical analysis},\ }\href@noop {} {\bibfield  {journal} {\bibinfo  {journal} {The American Naturalist}\ }\textbf {\bibinfo {volume} {184}},\ \bibinfo {pages} {439} (\bibinfo {year} {2014}{\natexlab{a}})}\BibitemShut {NoStop}%
\bibitem [{\citenamefont {Kalyuzhny}\ \emph {et~al.}(2014{\natexlab{b}})\citenamefont {Kalyuzhny}, \citenamefont {Schreiber}, \citenamefont {Chocron}, \citenamefont {Flather}, \citenamefont {Kadmon}, \citenamefont {Kessler},\ and\ \citenamefont {Shnerb}}]{kalyuzhny2014temporal}%
  \BibitemOpen
  \bibfield  {author} {\bibinfo {author} {\bibfnamefont {M.}~\bibnamefont {Kalyuzhny}}, \bibinfo {author} {\bibfnamefont {Y.}~\bibnamefont {Schreiber}}, \bibinfo {author} {\bibfnamefont {R.}~\bibnamefont {Chocron}}, \bibinfo {author} {\bibfnamefont {C.~H.}\ \bibnamefont {Flather}}, \bibinfo {author} {\bibfnamefont {R.}~\bibnamefont {Kadmon}}, \bibinfo {author} {\bibfnamefont {D.~A.}\ \bibnamefont {Kessler}},\ and\ \bibinfo {author} {\bibfnamefont {N.~M.}\ \bibnamefont {Shnerb}},\ }\bibfield  {title} {\bibinfo {title} {Temporal fluctuation scaling in populations and communities},\ }\href@noop {} {\bibfield  {journal} {\bibinfo  {journal} {Ecology}\ }\textbf {\bibinfo {volume} {95}},\ \bibinfo {pages} {1701} (\bibinfo {year} {2014}{\natexlab{b}})}\BibitemShut {NoStop}%
\bibitem [{\citenamefont {Chisholm}\ \emph {et~al.}(2014)\citenamefont {Chisholm}, \citenamefont {Condit}, \citenamefont {Rahman}, \citenamefont {Baker}, \citenamefont {Bunyavejchewin}, \citenamefont {Chen}, \citenamefont {Chuyong}, \citenamefont {Dattaraja}, \citenamefont {Davies}, \citenamefont {Ewango} \emph {et~al.}}]{chisholm2014temporal}%
  \BibitemOpen
  \bibfield  {author} {\bibinfo {author} {\bibfnamefont {R.~A.}\ \bibnamefont {Chisholm}}, \bibinfo {author} {\bibfnamefont {R.}~\bibnamefont {Condit}}, \bibinfo {author} {\bibfnamefont {K.~A.}\ \bibnamefont {Rahman}}, \bibinfo {author} {\bibfnamefont {P.~J.}\ \bibnamefont {Baker}}, \bibinfo {author} {\bibfnamefont {S.}~\bibnamefont {Bunyavejchewin}}, \bibinfo {author} {\bibfnamefont {Y.-Y.}\ \bibnamefont {Chen}}, \bibinfo {author} {\bibfnamefont {G.}~\bibnamefont {Chuyong}}, \bibinfo {author} {\bibfnamefont {H.}~\bibnamefont {Dattaraja}}, \bibinfo {author} {\bibfnamefont {S.}~\bibnamefont {Davies}}, \bibinfo {author} {\bibfnamefont {C.~E.}\ \bibnamefont {Ewango}}, \emph {et~al.},\ }\bibfield  {title} {\bibinfo {title} {Temporal variability of forest communities: empirical estimates of population change in 4000 tree species},\ }\href@noop {} {\bibfield  {journal} {\bibinfo  {journal} {Ecology Letters}\ }\textbf {\bibinfo {volume} {17}},\ \bibinfo {pages} {855} (\bibinfo {year} {2014})}\BibitemShut {NoStop}%
\bibitem [{\citenamefont {Kalyuzhny}\ \emph {et~al.}(2015)\citenamefont {Kalyuzhny}, \citenamefont {Kadmon},\ and\ \citenamefont {Shnerb}}]{kalyuzhny2015neutral}%
  \BibitemOpen
  \bibfield  {author} {\bibinfo {author} {\bibfnamefont {M.}~\bibnamefont {Kalyuzhny}}, \bibinfo {author} {\bibfnamefont {R.}~\bibnamefont {Kadmon}},\ and\ \bibinfo {author} {\bibfnamefont {N.~M.}\ \bibnamefont {Shnerb}},\ }\bibfield  {title} {\bibinfo {title} {A neutral theory with environmental stochasticity explains static and dynamic properties of ecological communities},\ }\href@noop {} {\bibfield  {journal} {\bibinfo  {journal} {Ecology Letters}\ }\textbf {\bibinfo {volume} {18}},\ \bibinfo {pages} {572} (\bibinfo {year} {2015})}\BibitemShut {NoStop}%
\bibitem [{\citenamefont {Chesson}\ and\ \citenamefont {Warner}(1981)}]{chesson1981environmental}%
  \BibitemOpen
  \bibfield  {author} {\bibinfo {author} {\bibfnamefont {P.~L.}\ \bibnamefont {Chesson}}\ and\ \bibinfo {author} {\bibfnamefont {R.~R.}\ \bibnamefont {Warner}},\ }\bibfield  {title} {\bibinfo {title} {Environmental variability promotes coexistence in lottery competitive systems},\ }\href@noop {} {\bibfield  {journal} {\bibinfo  {journal} {The American Naturalist}\ }\textbf {\bibinfo {volume} {117}},\ \bibinfo {pages} {923} (\bibinfo {year} {1981})}\BibitemShut {NoStop}%
\bibitem [{\citenamefont {Dean}\ \emph {et~al.}(2017)\citenamefont {Dean}, \citenamefont {Lehman},\ and\ \citenamefont {Yi}}]{dean2017fluctuating}%
  \BibitemOpen
  \bibfield  {author} {\bibinfo {author} {\bibfnamefont {A.~M.}\ \bibnamefont {Dean}}, \bibinfo {author} {\bibfnamefont {C.}~\bibnamefont {Lehman}},\ and\ \bibinfo {author} {\bibfnamefont {X.}~\bibnamefont {Yi}},\ }\bibfield  {title} {\bibinfo {title} {Fluctuating selection in the moran},\ }\href@noop {} {\bibfield  {journal} {\bibinfo  {journal} {Genetics}\ ,\ \bibinfo {pages} {genetics}} (\bibinfo {year} {2017})}\BibitemShut {NoStop}%
\bibitem [{\citenamefont {Danino}\ and\ \citenamefont {Shnerb}(2018)}]{danino2018theory}%
  \BibitemOpen
  \bibfield  {author} {\bibinfo {author} {\bibfnamefont {M.}~\bibnamefont {Danino}}\ and\ \bibinfo {author} {\bibfnamefont {N.~M.}\ \bibnamefont {Shnerb}},\ }\bibfield  {title} {\bibinfo {title} {Theory of time-averaged neutral dynamics with environmental stochasticity},\ }\href@noop {} {\bibfield  {journal} {\bibinfo  {journal} {Physical Review E}\ }\textbf {\bibinfo {volume} {97}},\ \bibinfo {pages} {042406} (\bibinfo {year} {2018})}\BibitemShut {NoStop}%
\bibitem [{\citenamefont {Pande}\ and\ \citenamefont {Shnerb}(2022)}]{pande2022temporal}%
  \BibitemOpen
  \bibfield  {author} {\bibinfo {author} {\bibfnamefont {J.}~\bibnamefont {Pande}}\ and\ \bibinfo {author} {\bibfnamefont {N.~M.}\ \bibnamefont {Shnerb}},\ }\bibfield  {title} {\bibinfo {title} {How temporal environmental stochasticity affects species richness: Destabilization, neutralization and the storage effect},\ }\href@noop {} {\bibfield  {journal} {\bibinfo  {journal} {Journal of Theoretical Biology}\ }\textbf {\bibinfo {volume} {539}},\ \bibinfo {pages} {111053} (\bibinfo {year} {2022})}\BibitemShut {NoStop}%
\bibitem [{\citenamefont {Meyer}\ \emph {et~al.}(2022)\citenamefont {Meyer}, \citenamefont {Steinmetz},\ and\ \citenamefont {Shnerb}}]{meyer2022storage}%
  \BibitemOpen
  \bibfield  {author} {\bibinfo {author} {\bibfnamefont {I.}~\bibnamefont {Meyer}}, \bibinfo {author} {\bibfnamefont {B.}~\bibnamefont {Steinmetz}},\ and\ \bibinfo {author} {\bibfnamefont {N.~M.}\ \bibnamefont {Shnerb}},\ }\bibfield  {title} {\bibinfo {title} {How the storage effect and the number of temporal niches affect biodiversity in stochastic and seasonal environments},\ }\href@noop {} {\bibfield  {journal} {\bibinfo  {journal} {PLOS Computational Biology}\ }\textbf {\bibinfo {volume} {18}},\ \bibinfo {pages} {e1009971} (\bibinfo {year} {2022})}\BibitemShut {NoStop}%
\bibitem [{\citenamefont {Dean}\ and\ \citenamefont {Shnerb}(2020)}]{dean2020stochasticity}%
  \BibitemOpen
  \bibfield  {author} {\bibinfo {author} {\bibfnamefont {A.}~\bibnamefont {Dean}}\ and\ \bibinfo {author} {\bibfnamefont {N.~M.}\ \bibnamefont {Shnerb}},\ }\bibfield  {title} {\bibinfo {title} {Stochasticity-induced stabilization in ecology and evolution: a new synthesis},\ }\href@noop {} {\bibfield  {journal} {\bibinfo  {journal} {Ecology}\ }\textbf {\bibinfo {volume} {101}},\ \bibinfo {pages} {e03098} (\bibinfo {year} {2020})}\BibitemShut {NoStop}%
\bibitem [{\citenamefont {Mallmin}\ \emph {et~al.}(2024)\citenamefont {Mallmin}, \citenamefont {Traulsen},\ and\ \citenamefont {De~Monte}}]{mallmin2024chaotic}%
  \BibitemOpen
  \bibfield  {author} {\bibinfo {author} {\bibfnamefont {E.}~\bibnamefont {Mallmin}}, \bibinfo {author} {\bibfnamefont {A.}~\bibnamefont {Traulsen}},\ and\ \bibinfo {author} {\bibfnamefont {S.}~\bibnamefont {De~Monte}},\ }\bibfield  {title} {\bibinfo {title} {Chaotic turnover of rare and abundant species in a strongly interacting model community},\ }\href@noop {} {\bibfield  {journal} {\bibinfo  {journal} {Proceedings of the National Academy of Sciences}\ }\textbf {\bibinfo {volume} {121}},\ \bibinfo {pages} {e2312822121} (\bibinfo {year} {2024})}\BibitemShut {NoStop}%
\bibitem [{\citenamefont {Steinmetz}\ \emph {et~al.}(2020)\citenamefont {Steinmetz}, \citenamefont {Kalyuzhny},\ and\ \citenamefont {Shnerb}}]{steinmetz2020intraspecific}%
  \BibitemOpen
  \bibfield  {author} {\bibinfo {author} {\bibfnamefont {B.}~\bibnamefont {Steinmetz}}, \bibinfo {author} {\bibfnamefont {M.}~\bibnamefont {Kalyuzhny}},\ and\ \bibinfo {author} {\bibfnamefont {N.~M.}\ \bibnamefont {Shnerb}},\ }\bibfield  {title} {\bibinfo {title} {Intraspecific variability in fluctuating environments: mechanisms of impact on species diversity},\ }\href@noop {} {\bibfield  {journal} {\bibinfo  {journal} {Ecology}\ }\textbf {\bibinfo {volume} {101}},\ \bibinfo {pages} {e03174} (\bibinfo {year} {2020})}\BibitemShut {NoStop}%
\bibitem [{\citenamefont {Maruvka}\ \emph {et~al.}(2010)\citenamefont {Maruvka}, \citenamefont {Shnerb},\ and\ \citenamefont {Kessler}}]{maruvka2010universal}%
  \BibitemOpen
  \bibfield  {author} {\bibinfo {author} {\bibfnamefont {Y.~E.}\ \bibnamefont {Maruvka}}, \bibinfo {author} {\bibfnamefont {N.~M.}\ \bibnamefont {Shnerb}},\ and\ \bibinfo {author} {\bibfnamefont {D.~A.}\ \bibnamefont {Kessler}},\ }\bibfield  {title} {\bibinfo {title} {Universal features of surname distribution in a subsample of a growing population},\ }\href@noop {} {\bibfield  {journal} {\bibinfo  {journal} {Journal of theoretical biology}\ }\textbf {\bibinfo {volume} {262}},\ \bibinfo {pages} {245} (\bibinfo {year} {2010})}\BibitemShut {NoStop}%
\bibitem [{\citenamefont {Ser-Giacomi}\ \emph {et~al.}(2018)\citenamefont {Ser-Giacomi}, \citenamefont {Zinger}, \citenamefont {Malviya}, \citenamefont {De~Vargas}, \citenamefont {Karsenti}, \citenamefont {Bowler},\ and\ \citenamefont {De~Monte}}]{ser2018ubiquitous}%
  \BibitemOpen
  \bibfield  {author} {\bibinfo {author} {\bibfnamefont {E.}~\bibnamefont {Ser-Giacomi}}, \bibinfo {author} {\bibfnamefont {L.}~\bibnamefont {Zinger}}, \bibinfo {author} {\bibfnamefont {S.}~\bibnamefont {Malviya}}, \bibinfo {author} {\bibfnamefont {C.}~\bibnamefont {De~Vargas}}, \bibinfo {author} {\bibfnamefont {E.}~\bibnamefont {Karsenti}}, \bibinfo {author} {\bibfnamefont {C.}~\bibnamefont {Bowler}},\ and\ \bibinfo {author} {\bibfnamefont {S.}~\bibnamefont {De~Monte}},\ }\bibfield  {title} {\bibinfo {title} {Ubiquitous abundance distribution of non-dominant plankton across the global ocean},\ }\href@noop {} {\bibfield  {journal} {\bibinfo  {journal} {Nature ecology \& evolution}\ }\textbf {\bibinfo {volume} {2}},\ \bibinfo {pages} {1243} (\bibinfo {year} {2018})}\BibitemShut {NoStop}%
\bibitem [{\citenamefont {Arnoulx~de Pirey}\ and\ \citenamefont {Bunin}(2024)}]{arnoulx2024many}%
  \BibitemOpen
  \bibfield  {author} {\bibinfo {author} {\bibfnamefont {T.}~\bibnamefont {Arnoulx~de Pirey}}\ and\ \bibinfo {author} {\bibfnamefont {G.}~\bibnamefont {Bunin}},\ }\bibfield  {title} {\bibinfo {title} {Many-species ecological fluctuations as a jump process from the brink of extinction},\ }\href@noop {} {\bibfield  {journal} {\bibinfo  {journal} {Physical Review X}\ }\textbf {\bibinfo {volume} {14}},\ \bibinfo {pages} {011037} (\bibinfo {year} {2024})}\BibitemShut {NoStop}%
\bibitem [{\citenamefont {Hutchinson}(1961{\natexlab{b}})}]{hutchinson1961paradox}%
  \BibitemOpen
  \bibfield  {author} {\bibinfo {author} {\bibfnamefont {G.~E.}\ \bibnamefont {Hutchinson}},\ }\bibfield  {title} {\bibinfo {title} {The paradox of the plankton},\ }\href@noop {} {\bibfield  {journal} {\bibinfo  {journal} {The American Naturalist}\ ,\ \bibinfo {pages} {137}} (\bibinfo {year} {1961}{\natexlab{b}})}\BibitemShut {NoStop}%
\bibitem [{\citenamefont {Lande}\ \emph {et~al.}(2003)\citenamefont {Lande}, \citenamefont {Engen},\ and\ \citenamefont {Saether}}]{lande2003stochastic}%
  \BibitemOpen
  \bibfield  {author} {\bibinfo {author} {\bibfnamefont {R.}~\bibnamefont {Lande}}, \bibinfo {author} {\bibfnamefont {S.}~\bibnamefont {Engen}},\ and\ \bibinfo {author} {\bibfnamefont {B.-E.}\ \bibnamefont {Saether}},\ }\href@noop {} {\emph {\bibinfo {title} {Stochastic population dynamics in ecology and conservation}}}\ (\bibinfo  {publisher} {Oxford University Press},\ \bibinfo {year} {2003})\BibitemShut {NoStop}%
\bibitem [{\citenamefont {Hekstra}\ and\ \citenamefont {Leibler}(2012)}]{hekstra2012contingency}%
  \BibitemOpen
  \bibfield  {author} {\bibinfo {author} {\bibfnamefont {D.~R.}\ \bibnamefont {Hekstra}}\ and\ \bibinfo {author} {\bibfnamefont {S.}~\bibnamefont {Leibler}},\ }\bibfield  {title} {\bibinfo {title} {Contingency and statistical laws in replicate microbial closed ecosystems},\ }\href@noop {} {\bibfield  {journal} {\bibinfo  {journal} {Cell}\ }\textbf {\bibinfo {volume} {149}},\ \bibinfo {pages} {1164} (\bibinfo {year} {2012})}\BibitemShut {NoStop}%
\bibitem [{\citenamefont {Batista-Tom{\'a}s}\ \emph {et~al.}(2021)\citenamefont {Batista-Tom{\'a}s}, \citenamefont {De~Martino},\ and\ \citenamefont {Mulet}}]{batista2021path}%
  \BibitemOpen
  \bibfield  {author} {\bibinfo {author} {\bibfnamefont {A.}~\bibnamefont {Batista-Tom{\'a}s}}, \bibinfo {author} {\bibfnamefont {A.}~\bibnamefont {De~Martino}},\ and\ \bibinfo {author} {\bibfnamefont {R.}~\bibnamefont {Mulet}},\ }\bibfield  {title} {\bibinfo {title} {Path-integral solution of macarthur’s resource-competition model for large ecosystems with random species-resources couplings},\ }\href@noop {} {\bibfield  {journal} {\bibinfo  {journal} {Chaos: An Interdisciplinary Journal of Nonlinear Science}\ }\textbf {\bibinfo {volume} {31}} (\bibinfo {year} {2021})}\BibitemShut {NoStop}%
\bibitem [{\citenamefont {Waide}\ \emph {et~al.}(1999)\citenamefont {Waide}, \citenamefont {Willig}, \citenamefont {Steiner}, \citenamefont {Mittelbach}, \citenamefont {Gough}, \citenamefont {Dodson}, \citenamefont {Juday},\ and\ \citenamefont {Parmenter}}]{waide1999relationship}%
  \BibitemOpen
  \bibfield  {author} {\bibinfo {author} {\bibfnamefont {R.}~\bibnamefont {Waide}}, \bibinfo {author} {\bibfnamefont {M.}~\bibnamefont {Willig}}, \bibinfo {author} {\bibfnamefont {C.}~\bibnamefont {Steiner}}, \bibinfo {author} {\bibfnamefont {G.}~\bibnamefont {Mittelbach}}, \bibinfo {author} {\bibfnamefont {L.}~\bibnamefont {Gough}}, \bibinfo {author} {\bibfnamefont {S.}~\bibnamefont {Dodson}}, \bibinfo {author} {\bibfnamefont {G.}~\bibnamefont {Juday}},\ and\ \bibinfo {author} {\bibfnamefont {R.}~\bibnamefont {Parmenter}},\ }\bibfield  {title} {\bibinfo {title} {The relationship between productivity and species richness},\ }\href@noop {} {\bibfield  {journal} {\bibinfo  {journal} {Annual review of Ecology and Systematics}\ }\textbf {\bibinfo {volume} {30}},\ \bibinfo {pages} {257} (\bibinfo {year} {1999})}\BibitemShut {NoStop}%
\bibitem [{\citenamefont {Kadmon}\ and\ \citenamefont {Benjamini}(2006)}]{kadmon2006effects}%
  \BibitemOpen
  \bibfield  {author} {\bibinfo {author} {\bibfnamefont {R.}~\bibnamefont {Kadmon}}\ and\ \bibinfo {author} {\bibfnamefont {Y.}~\bibnamefont {Benjamini}},\ }\bibfield  {title} {\bibinfo {title} {Effects of productivity and disturbance on species richness: a neutral model},\ }\href@noop {} {\bibfield  {journal} {\bibinfo  {journal} {The American Naturalist}\ }\textbf {\bibinfo {volume} {167}},\ \bibinfo {pages} {939} (\bibinfo {year} {2006})}\BibitemShut {NoStop}%
\bibitem [{\citenamefont {Loreau}\ and\ \citenamefont {de~Mazancourt}(2008)}]{loreau2008species}%
  \BibitemOpen
  \bibfield  {author} {\bibinfo {author} {\bibfnamefont {M.}~\bibnamefont {Loreau}}\ and\ \bibinfo {author} {\bibfnamefont {C.}~\bibnamefont {de~Mazancourt}},\ }\bibfield  {title} {\bibinfo {title} {Species synchrony and its drivers: neutral and nonneutral community dynamics in fluctuating environments},\ }\href@noop {} {\bibfield  {journal} {\bibinfo  {journal} {The American Naturalist}\ }\textbf {\bibinfo {volume} {172}},\ \bibinfo {pages} {E48} (\bibinfo {year} {2008})}\BibitemShut {NoStop}%
\bibitem [{\citenamefont {Chesson}(1982)}]{chesson1982storage}%
  \BibitemOpen
  \bibfield  {author} {\bibinfo {author} {\bibfnamefont {P.~L.}\ \bibnamefont {Chesson}},\ }\bibfield  {title} {\bibinfo {title} {The storage effect in stochastic competition models},\ }\href@noop {} {\bibfield  {journal} {\bibinfo  {journal} {Mathematical ecology: Proceedings, Trieste}\ ,\ \bibinfo {pages} {76}} (\bibinfo {year} {1982})}\BibitemShut {NoStop}%
\bibitem [{\citenamefont {Ellner}\ \emph {et~al.}(2016)\citenamefont {Ellner}, \citenamefont {Snyder},\ and\ \citenamefont {Adler}}]{ellner2016quantify}%
  \BibitemOpen
  \bibfield  {author} {\bibinfo {author} {\bibfnamefont {S.~P.}\ \bibnamefont {Ellner}}, \bibinfo {author} {\bibfnamefont {R.~E.}\ \bibnamefont {Snyder}},\ and\ \bibinfo {author} {\bibfnamefont {P.~B.}\ \bibnamefont {Adler}},\ }\bibfield  {title} {\bibinfo {title} {How to quantify the temporal storage effect using simulations instead of math},\ }\href@noop {} {\bibfield  {journal} {\bibinfo  {journal} {Ecology Letters}\ }\textbf {\bibinfo {volume} {19}},\ \bibinfo {pages} {1333} (\bibinfo {year} {2016})}\BibitemShut {NoStop}%
\bibitem [{\citenamefont {Letten}\ \emph {et~al.}(2018)\citenamefont {Letten}, \citenamefont {Dhami}, \citenamefont {Ke},\ and\ \citenamefont {Fukami}}]{letten2018species}%
  \BibitemOpen
  \bibfield  {author} {\bibinfo {author} {\bibfnamefont {A.~D.}\ \bibnamefont {Letten}}, \bibinfo {author} {\bibfnamefont {M.~K.}\ \bibnamefont {Dhami}}, \bibinfo {author} {\bibfnamefont {P.-J.}\ \bibnamefont {Ke}},\ and\ \bibinfo {author} {\bibfnamefont {T.}~\bibnamefont {Fukami}},\ }\bibfield  {title} {\bibinfo {title} {Species coexistence through simultaneous fluctuation-dependent mechanisms},\ }\href@noop {} {\bibfield  {journal} {\bibinfo  {journal} {Proceedings of the National Academy of Sciences}\ }\textbf {\bibinfo {volume} {115}},\ \bibinfo {pages} {6745} (\bibinfo {year} {2018})}\BibitemShut {NoStop}%
\bibitem [{\citenamefont {Chesson}\ and\ \citenamefont {Huntly}(1989)}]{chesson1989short}%
  \BibitemOpen
  \bibfield  {author} {\bibinfo {author} {\bibfnamefont {P.~L.}\ \bibnamefont {Chesson}}\ and\ \bibinfo {author} {\bibfnamefont {N.}~\bibnamefont {Huntly}},\ }\bibfield  {title} {\bibinfo {title} {Short-term instabilities and long-term community dynamics},\ }\href@noop {} {\bibfield  {journal} {\bibinfo  {journal} {Trends in Ecology \& Evolution}\ }\textbf {\bibinfo {volume} {4}},\ \bibinfo {pages} {293} (\bibinfo {year} {1989})}\BibitemShut {NoStop}%
\bibitem [{\citenamefont {Holt}(2006)}]{holt2006emergent}%
  \BibitemOpen
  \bibfield  {author} {\bibinfo {author} {\bibfnamefont {R.~D.}\ \bibnamefont {Holt}},\ }\bibfield  {title} {\bibinfo {title} {Emergent neutrality},\ }\href@noop {} {\bibfield  {journal} {\bibinfo  {journal} {Trends in ecology \& evolution}\ }\textbf {\bibinfo {volume} {21}},\ \bibinfo {pages} {531} (\bibinfo {year} {2006})}\BibitemShut {NoStop}%
\bibitem [{\citenamefont {Vergnon}\ \emph {et~al.}(2012)\citenamefont {Vergnon}, \citenamefont {Van~Nes},\ and\ \citenamefont {Scheffer}}]{vergnon2012emergent}%
  \BibitemOpen
  \bibfield  {author} {\bibinfo {author} {\bibfnamefont {R.}~\bibnamefont {Vergnon}}, \bibinfo {author} {\bibfnamefont {E.~H.}\ \bibnamefont {Van~Nes}},\ and\ \bibinfo {author} {\bibfnamefont {M.}~\bibnamefont {Scheffer}},\ }\bibfield  {title} {\bibinfo {title} {Emergent neutrality leads to multimodal species abundance distributions},\ }\href@noop {} {\bibfield  {journal} {\bibinfo  {journal} {Nature communications}\ }\textbf {\bibinfo {volume} {3}},\ \bibinfo {pages} {663} (\bibinfo {year} {2012})}\BibitemShut {NoStop}%
\bibitem [{\citenamefont {Roy}\ \emph {et~al.}(2019)\citenamefont {Roy}, \citenamefont {Biroli}, \citenamefont {Bunin},\ and\ \citenamefont {Cammarota}}]{roy2019numerical}%
  \BibitemOpen
  \bibfield  {author} {\bibinfo {author} {\bibfnamefont {F.}~\bibnamefont {Roy}}, \bibinfo {author} {\bibfnamefont {G.}~\bibnamefont {Biroli}}, \bibinfo {author} {\bibfnamefont {G.}~\bibnamefont {Bunin}},\ and\ \bibinfo {author} {\bibfnamefont {C.}~\bibnamefont {Cammarota}},\ }\bibfield  {title} {\bibinfo {title} {Numerical implementation of dynamical mean field theory for disordered systems: Application to the lotka--volterra model of ecosystems},\ }\href@noop {} {\bibfield  {journal} {\bibinfo  {journal} {Journal of Physics A: Mathematical and Theoretical}\ }\textbf {\bibinfo {volume} {52}},\ \bibinfo {pages} {484001} (\bibinfo {year} {2019})}\BibitemShut {NoStop}%
\end{thebibliography}%

\clearpage
\renewcommand{\appendixesname}{Supporting Information}
\renewcommand{\appendixname}{SI}
\appendix

\begin{center}
    \bf{Supporting Information}

\end{center}

\section{The necessity of immigration} \label{suppC}

\citet{van2024tiny} have considered various models were proposed for a diverse community in which all species have the same mean fitness. Most cases were reviewed in the supplemental material of the aforementioned article, but the version analyzed in detail in the main text is,
\begin{equation} \label{eqS7}
   \frac{ dN_i}{dt}  =  N_i \left( 1-\frac{\sum_{j=1}^S N_j}{K} \right) dt +  {\tilde \sigma}_e \eta_i(t) N_i, 
\end{equation}
where $\eta(t)$ is a white noise process. Ito calculus is applied, so the numerical integration procedure satisfies,  
\begin{equation} \label{eqS8}
N_{i,t+\Delta t} = N_{i,t} + \Delta t \left[\mu + N_{i,t} \left( 1-\frac{\sum_{j=1}^S N_{j,t}}{K}   \right) \right] + {\tilde \sigma_e} \sqrt{\Delta t} {\cal N}(0,1) N_{i,t},
\end{equation}
where ${\cal N}(0,1)$ is a random variable drawn (independently for each $\Delta t$ and each species) from a normal distribution with zero mean and unit variance.

\begin{figure}[hbt!]
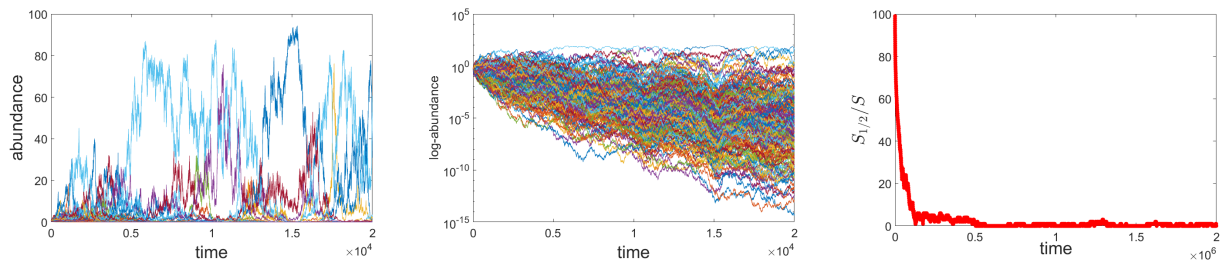

	\centering{
		\includegraphics[width=5.5cm]{figSupp_a.png}  	\includegraphics[width=5.5cm]{figSupp_b.png} \includegraphics[width=5.5cm]{figSupp_c.png}}
	\caption{Species abundances (left), log-abundance (middle) and the evenness parameter $S_{1/2}/S$ (right) for the model considered in the main text of~\cite{van2024tiny}, Parameters are $S=100$, ${\tilde \sigma_e} = 0.005$ and $K = 100$. The dynamic of  Eq. (\ref{eqS8}) was integrated. The log-abundance of all species performs a random walk (middle), hence, as time goes by, more and more species get stuck at negligible abundances (this is the diffusive trapping, or stickiness of the rare state, left), and only a single species dominates the community  and the evenness parameter decreases to  $1/S$     \label{figS7}}
\end{figure}

However, in the absence of a immigration term, i.e., when  $\mu=0$, such a process leads to monodominance, the state where only one species survives and the abundance of all other species drop below any finite value. Fig. \ref{figS7} we demonstrate numerically this behavior.

The results presented in our main text explain this phenomenon. First, if  $\mu=0$ then $\alpha=0$ in our solution, Eq. (\ref{eq5}). The resulting species abundance distribution  diverges at zero abundance and cannot be normalized. The interpretation of such non-normalizable abundance distribution is that the abundance of all species will drop below any finite threshold as time goes to infinity.

One may arrive at the same conclusion from a different perspective~\cite{dean2020stochasticity}. Let's assume we have two species in a community with a fixed and rigid carrying capacity (i.e., they are playing a zero-sum game) and environmental stochasticity that determines their relative fitness.

Denoting by $x$ the fraction of the focal species, its dynamics satisfies $dx/dt = s(t) x(1-x)$, where $s(t)$ reflects the effect of the fluctuating environment. Therefore, the population undergoes a balanced walk in the logit space $z=\ln[x/(1-x)]$. In this type of random walk, a species that ``gets stuck'' in the low-density region will remain there for increasing periods, as described in Section 5 of~\cite{dean2020stochasticity}. This is the ``diffusive trapping'', or the stickiness of low-abundance states, that causes the system to be dominated at any given moment by a single species.

Now we can easily generalize this argument for $S$ species: the relative fitness of each species (i.e., its fitness relative to the average fitness of the community) is a zero-mean random variable. Therefore, in the log-abundance space, it undergoes a random walk, and so all species except one will fall below any threshold of abundance over time. Thus, as seen in Figure \ref{figS7}, the system in the long run is dominated by a single species, and the abundance of all other species is negligible.

\section{Derivation of Eq. (\ref{eq5})} \label{suppD}

The Fokker-Planck equation that corresponds to Eq. (\ref{eq4}) is 
\begin{equation} \label{eqS10}
  \frac{\partial P(n,t)}{\partial n} = \frac{\sigma^2}{2} \frac{\partial^2}{\partial n^2} \left( n^2 P(n) \right) + \left( d-\frac{\sigma^2}{2} \right) \frac{\partial}{\partial n} \left( n P(n) \right) - \mu \frac{\partial P(n)}{\partial n}
\end{equation}
In steady state, $P(n)$ is time-independent and so its time derivative vanishes, leaving one  to solve the exact ordinary differential equation (after first integration),
\begin{equation} 
\frac{\sigma^2}{2} \frac{\partial}{\partial n} \left( n^2 P(n) \right) + \left( d-\frac{\sigma^2}{2} \right)  \left( n P(n) \right) - \mu P(n) = 0.
\end{equation}
Here we have put the integration constant to zero to avoid diverging solutions that yield non-normalizable $P(n)$ functions. Therefore, 
\begin{equation} 
\frac{\sigma^2}{2} n^2  \frac{\partial P(n)}{\partial n} +\left( d+\frac{\sigma^2}{2} \right) n  P  - \mu P(n) = 0.
\end{equation}
That yields Eq. (\ref{eq5}) of the main text.

\section{From egalitarian community to hyperdominant species} \label{suppE}
 
Let us consider a community of $J$ individuals, whose species richness is $S$. By definition,  
\begin{equation} \label{eqS1}
S = \int_0^\infty  P(n) dn, 
\end{equation}
and 
\begin{equation} \label{eqS2}
J = \int_0^\infty n P(n) dn,
\end{equation}

From Eq. (\ref{eq5}) of the main text we take $P(n)$ and plug it into Eq. (\ref{eqS1}). First we implement the condition on $S$ to determine the normalization factor  $A$, and find, 
\begin{equation}
    P(n) = S \frac{\alpha^{\beta-1} }{\Gamma[\beta-1] }e^{-\alpha/n} n^{-\beta}
\end{equation}

Now let us assume $\beta>2$. In that case the integral in (\ref{eqS2}) converges so that,
\begin{equation} \label{eqS3}
    J = \frac{\alpha}{\beta-2} S. 
\end{equation}

As a metric for the evenness of the community, let us take the one used in \cite{van2024tiny}. First, we will check the richness $S_{1/2}$ of the minimum set of species required to contain at least fifty percent of the entire community population. If the species within this set with the smallest abundance  is represented by $n_{1/2}$ individuals, 
this implies (implementing Eq. (\ref{eqS3})
\begin{equation}
    \int_{n_{1/2}}^\infty nP(n) dn = \alpha  S \left( (\beta -1) - \frac{\Gamma[\beta-2,\alpha/n_{1/2}]}{\Gamma[\beta -1]} \right) = J/2 = \frac{\alpha}{2(\beta-2)} S,
\end{equation}

When $\mu \to 0$ then $\alpha \to 0$  and hence, 
\begin{equation}
    \Gamma[\beta-2,\alpha/n_{1/2}] \approx \Gamma[\beta-2]- \frac{1}{\beta-2} \left(\frac{\alpha}{n_{1/2}}\right)^{\beta-2}
\end{equation}
therefore
\begin{equation}
 \frac{\alpha}{n_{1/2}} = \left( \frac{\Gamma[\beta-1]}{2} \right)^{1/(\beta-2)},
\end{equation}
so the ratio $\alpha/n_{1/2}$ approaches zero as $\beta \to 2^+$

The fraction of species whose abundance is equal or greater than $n_{1/2}$ is obtained as 
\begin{equation}
   \frac{ S_{1/2}}{S}  = \frac{1}{S}  \int_{n_{1/2}}^\infty P(n) dn = 1- \frac{\Gamma[\beta-1,\alpha/n_{1/2}]}{\Gamma[\beta-1]} = \frac{1}{\Gamma[\beta]} \left(\frac{\Gamma[\beta-1]}{2}\right)^{\frac{\beta-1}{\beta-2}}.
\end{equation}
In particular, as $\beta \to 2^+$,
\begin{equation}
     \frac{S_{1/2}}{S} \sim e^{-(\ln 2)/(\beta-2)}. 
\end{equation}

\section{Demographic stochasticity} \label{SuppE}

Every group of living organisms is influenced by various types of stochasticity. The literature~\cite{lande2003stochastic,kalyuzhny2014niche}, usually distinguishes between {\it environmental stochasticity}, which affects populations coherently, and {\it demographic stochasticity}, which affects the fitness of individuals or small groups of individuals in an incoherent manner, so that the overall reproductive success of the population reflects the sum of many random events. Weather (such as temperature or precipitation)  fluctuations that are correlated over a large area and simultaneously affect  entire populations contribute to environmental stochasticity, while the degree of success or failure of individuals in finding food or avoiding predators relative to its population mean contribute to demographic stochasticity.

Environmental stochasticity is represented in equations by noise that is proportional to the population size. Thus, in equation \ref{eq2} of the main text, the term $r_i(t)$, which varies stochastically, multiplies $N_i$. This mathematical expression reflects the situation where, in a good year, the average number of offspring for each individual increases, and in a bad year, it decreases. Demographic stochasticity, on the other hand, represents the success or failure of individuals, so that the overall result, according to the Central Limit Theorem, is fluctuations that are proportional to the square root of the population size. The corresponding term in the Langevin equation is $\eta(t)\sqrt{N_i}$. 

In a Fokker-Planck equation, the term that represents demographic stoahcsticity is
\begin{equation}
\frac{\sigma_d^2}{2} \frac{\partial^2}{\partial n^2} \left( n P(n) \right) 
\end{equation}
We denote the strength of demographic stochasticity (i.e., the variance in the number of offspring per individual) by $\sigma_d^2$, distinguishing it from $\sigma^2$, which represents the strength of environmental stochasticity.

The extended version of Eq. (\ref{eqS10})  thus takes the form,
\begin{equation} \label{eqS11}
  \frac{\partial P(n,t)}{\partial n} = \frac{\sigma^2}{2} \frac{\partial^2}{\partial n^2} \left( n^2 P(n) \right) +\frac{\sigma_d^2}{2} \frac{\partial^2}{\partial n^2} \left( n P(n) \right) + \left( d-\frac{\sigma^2}{2} \right) \frac{\partial}{\partial n} \left( n P(n) \right) - \mu \frac{\partial P(n)}{\partial n}
\end{equation}

Again, in steady state, $P(n)$ is time-independent,
\begin{equation} 
\frac{\sigma^2}{2} \frac{\partial}{\partial n} \left( n^2 P(n) \right) + \frac{\sigma_d^2}{2} \frac{\partial}{\partial n} \left( n P(n) \right) + \left( d-\frac{\sigma^2}{2} \right)  \left( n P(n) \right) - \mu P(n) = 0,
\end{equation}
and therfore the species abundance distribution is given by
\begin{equation} \label{eqS12}
P(n) = A n^{-1-2\mu/\sigma_d^2} \left(\frac{\sigma_d^2}{\sigma^2} + n\right)^{-2d/\sigma^2 - 2\mu/\sigma_d^2}   
\end{equation}

\end{document}